\newcommand \Mpc {h^{-1}{\rm Mpc}}

\newcommand \arcm{\hbox{$^{\prime}$}}
\newcommand \arcs{\hbox{$^{\prime\prime}$}}

\newcommand \kms {{\rm km~s}^{-1}}
\newcommand \msun {h^{-1} M_\odot}
\newcommand \beqn {\begin{equation}}
\newcommand \eeqn {\end{equation}}
\newcommand \ncirs {72 }

\documentclass[12pt,preprint]{aastex}
\usepackage{emulateapj5}

\begin{document}

\title{The Virial Mass Function of Nearby SDSS Galaxy Clusters}

\author{Kenneth Rines\altaffilmark{1,2},
Antonaldo Diaferio\altaffilmark{3}, 
and Priyamvada Natarajan\altaffilmark{1,4}} 
\email{krines@astro.yale.edu}

\altaffiltext{1}{Yale Center for Astronomy and Astrophysics, Yale University, PO Box 208121, New Haven, CT 06520-8121; krines@astro.yale.edu}
\altaffiltext{2}{Current Address: Smithsonian Astrophysical Observatory, 60 Garden St, MS 20, Cambridge, MA 02138; krines@cfa.harvard.edu}
\altaffiltext{3}{Universit\`a degli Studi di Torino,
Dipartimento di Fisica Generale ``Amedeo Avogadro'', Torino, Italy; diaferio@ph.unito.it}
\altaffiltext{4}{Department of Astronomy, Yale University, PO Box 208121, New Haven, CT 06520-8121; priya@astro.yale.edu}

\begin{abstract}

We present a new determination of the cluster mass function and
velocity dispersion function in a volume $\sim$10$^7 h^3$Mpc$^{-3}$
using data from the Fourth Data Release of the Sloan Digital Sky
Survey (SDSS) to determine virial masses.  We use the caustic
technique to remove foreground and background galaxies.  The cluster
virial mass function agrees well with recent estimates from both
X-ray observations and cluster richnesses.  Our determination of the
mass function lies between those predicted by the First-Year and
Three-Year WMAP data.  We constrain the cosmological parameters
$\Omega_m$ and $\sigma_8$ and find good agreement with WMAP and
constraints from other techniques.  With the CIRS mass function alone,
we estimate $\Omega_m=0.24^{+0.14}_{-0.09}$ and
$\sigma_8=0.92^{+0.24}_{-0.19}$, or $\sigma_8=0.84\pm$0.03 when
holding $\Omega_m=0.3$ fixed.  We also use the WMAP parameters as
priors and constrain velocity segregation in clusters.  Using the
First and Third-Year results, we infer velocity segregation of
$\sigma_{gxy}/\sigma_{DM}\approx$0.94$\pm$0.05 or 1.28$\pm$0.06
respectively.  The good agreement of various estimates of the cluster
mass function shows that it is a useful independent constraint on
estimates of cosmological parameters.  We compare the velocity
dispersion function of clusters to that of early-type galaxies and
conclude that clusters comprise the high-velocity end of the velocity
dispersion function of dark matter haloes.  Future studies of galaxy
groups are needed to study the transition between dark matter haloes
containing individiual galaxies and those containing systems of
galaxies.  The evolution of cluster abundances provides constraints on
dark energy models; the mass function presented here offers an
important low redshift calibration benchmark.

\end{abstract}


\section{Introduction}

Clusters of galaxies are the most massive gravitationally relaxed
systems in the universe, so the observed cluster mass function is a
sensitive probe of cosmological parameters.  Galaxy cluster abundances
are most sensitive to the matter density of the universe $\Omega_m$
and $\sigma_8$, the rms fluctuations in spheres of radius 8$\Mpc$ and
the usual normalization of the linear power spectrum
\citep{henry91,bahcall93}.  The evolution of the mass function is an
excellent probe of the growth of structure, and thus can constrain the
properties of dark energy
\citep[e.g.,][]{haiman01,hu03b,vikhlinin03,henry04,majumdar04,vv04}.  Robust
constraints from cluster evolution are only possible with an accurate
determination of the cluster mass function in the nearby universe.

Most recent estimates of $\sigma_8$ from the local cluster mass function
find values of $\sigma_8$$\approx$0.6-0.8 for $\Omega_m$=0.3
\citep{pierpaoli01,hiflugcs,seljak02,viana02,pierpaoli03,schuecker03,viana03,bahcall03a,vv04,eke05,dahle06} (there is a strong
degeneracy between $\sigma_8$ and $\Omega_m$).  Estimates based on
cosmic shear usually find larger values of $\sigma_8$$\approx$0.8-1.0
\citep{massey05,hoekstra06,semboloni06}, although some investigators 
find smaller values \citep{hamana03,heymans05}.  Recent attempts to
model the relation between dark matter haloes and galaxies
\citep{2003MNRAS.345..923V,tinker05} suggest that smaller values of
$\sigma_8$$\approx$0.6-0.8 are required to match cluster mass-to-light
ratios \citep{cnoc96,bahcall2000,cairnsii}.  Various estimates based
on bulk flow models seem to be converging on intermediate values
\citep[see Table 3 of][]{pike05}.  The WMAP satellite has produced tight
constraints on many cosmological parameters
\citep{spergel03,spergel06}.  Among the largest changes between the
First-Year and Three-Year WMAP results (hereafter WMAP1 and WMAP3) was
a decrease in both $\sigma_8$ and $\Omega_m$
\citep{spergel06}.  The WMAP3 results provide a much better
fit to the mass function of X-ray clusters from the ROSAT HIghest
X-ray FLUx Galaxy Cluster Survey \citep[HIFLUGCS,][]{hiflugcs} than the
WMAP1 results \citep{reiprich06}.

The greatest difficulty in measuring the cluster mass function is
obtaining sufficiently accurate mass estimates.  Three
well-established methods exist for measuring cluster masses.  The
dynamics of cluster galaxies provided the first evidence for dark
matter when
\citet{zwicky1933,zwicky1937} applied the virial theorem to the Coma
cluster.  The second method is to use the properties of the hot
intracluster medium (ICM), whose distribution and temperature probe
the gravitational potential of the cluster.  Finally, gravitational
lensing can provide very accurate mass estimates in the centers of
clusters and somewhat less accurate estimates at large radii.

All three techniques have been used extensively, but they are each
subject to potentially large systematic uncertainties.  There may be
velocity bias between cluster galaxies and dark matter particles,
although the magnitude and direction of this bias is under debate
\citep[e.g.,][]{kauffmann1999b,diemand04,faltenbacher05,benatov06}.  {\em
Chandra} and {\em XMM-Newton} observations show that the physical
properties of the ICM are complex and involve significant interaction
with AGN
\citep[e.g.,][]{2000ApJ...534L.135M,2000ApJ...541..542M,fabian00,2001ApJ...551..160V,fabian06}.
Finally, mass estimates from gravitational lensing (especially at
large radii) are subject to significant uncertainties due to lensing
by other structures along the line-of-sight 
\citep[e.g.,][]{2001ApJ...547..560M}, although these might be overcome 
by combining information from strong and weak lensing
\citep{natarajan02,bradac04}.  These potential systematics have led
some investigators to invert the problem and determine cluster
properties by matching the mass function predicted by a cosmological
model to the observed richness function or luminosity function
\citep{berlind06,stanek06}.

Despite these challenges, the cluster mass function is the subject of
much study.  Most recent measurements of the cluster mass function
have been made with either X-ray data \citep{hiflugcs,allen03,vv04} or
optical richness data \citep{bahcall03a}.  The most significant
uncertainty in determining $\sigma_8$ from X-ray observations is the
normalization of scaling relations, either the $M-T_X$ or the $M-L_X$
relation.  The normalizations are often determined from hydrodynamical
simulations \citep[e.g.,][]{borgani01}, but different simulations
produce a range of normalizations \citep{pierpaoli03,viana03}.
Another potential concern in determining scaling relations is
Malmquist bias \citep{stanek06}.  \citet{vv04} avoid these scaling
relations by estimating the mass of the intracluster medium (ICM) and
measuring the baryonic mass function \citep[a similar analysis was
presented by][]{shimasaku97}.  This method requires assumptions about:
the relative contribution of stars and gas to the total baryon mass,
the ratio of the baryon fraction in clusters to the global value, and
the mass dependence of this ratio.  A potentially large systematic
uncertainty in all of these X-ray studies is a possible offset between
the temperature $T_{spec}$ measured by X-ray satellites and the
emission-weighted temperature $T_{ew}$ calculated in hydrodynamical
simulations \citep{rasia05}.  This effect can be understood as the
excess contribution of line emission from cooler gas in an ICM with a
variety of temperatures, thus causing $T_{spec}$ to be an
underestimate of $T_{ew}$.  Correcting for this effect would increase
the estimated cluster masses and thus increase the best-fit values of
$\sigma_8$ \citep{rasia05,shimizu06}.  A similar systematic effect is
suggested by the higher normalization of the mass-temperature relation
found by \citet{vikhlinin06} from a study of relaxed clusters with
temperature profiles from {\em Chandra} data.  An increase in this
normalization would increase the inferred cluster masses and hence the
overall mass function, shifting the constraints towards higher values
of $\Omega_m$ and $\sigma_8$.

\citet{viana02} use a hybrid approach of combining the X-ray
luminosity function of clusters with a mass-luminosity relation
determined from weak lensing \citep{sheldon01}.  They find results
consistent with other X-ray studies, but note that large systematic
uncertainties are possible.  Recently, \citet{dahle06} used weak
lensing mass estimates of an X-ray selected cluster sample to measure
the mass function.  He found good agreement with many of the
constraints from X-ray estimates but lower values of $\Omega_m$ and
$\sigma_8$ than recent cosmic shear studies.

To our knowledge, the most recent cluster virial mass function is that
of \citet{girardi98b}.  The advent of large-scale redshift surveys,
especially the Sloan Digital Sky Survey \citep[SDSS,][]{sdss}, enables
more accurate virial mass estimates due to their size and uniformity.
Compared to X-ray studies, virial mass estimates are sensitive to
larger scales ($r_{200}$ rather than $r_{500}$, where $r_{\Delta}$ is
the radius within which the enclosed density is $\Delta$ times the
critical density), which allow for comparison with theoretical mass
functions with significantly less extrapolation \citep{white02}.
Also, virial masses can be estimated for poor clusters and rich
groups, whereas X-ray mass estimates of these systems are complicated
by possible energy input from supernovae and AGN
\citep[e.g.,][]{loewenstein00}.  Probing these smaller masses enables
a direct constraint on fluctuations on the scale 8$\Mpc$, rather than
the $\sim$14$\Mpc$ scales probed by $\sim$10$^{15}\msun$ clusters
\citep{pierpaoli01}. \citet{eke05} used a simplified version of the
virial theorem to estimate the group mass function from an
optically-selected group catalog in the Two Degree Field Galaxy
Redshift Survey \citep{2df}.  Differences in our analysis include:
X-ray selection rather than optical selection of clusters, use of the
full virial theorem with corrections for the surface pressure term,
much better sampling of individual systems (the average number of
cluster members is 50), and more conservative rejection of interlopers
(thus significantly reducing scatter in the mass estimates).
Estimates of the group mass function using virial masses are usually
limited by systematics in group selection, mass estimation, and cosmic
variance \citep{girardi00b,martinez02b,heinamaki03}.

Here we use the virial theorem to estimate cluster masses of an X-ray
selected sample of clusters with data from the Sloan Digital Sky
Survey.  The use of X-ray selection reduces the impact of projection
effects on cluster selection.  Perhaps the greatest advantage of X-ray
selection of our sample is that the selection function is well
understood and can be computed directly.  The cluster sample we use is
the Cluster Infall Regions in SDSS (CIRS) sample of \citet{cirsi} with
some minor modifications described below.

We describe the data and the cluster sample in $\S$ 2.  In $\S$ 3, we
estimate the mass function using both virial masses and caustic masses
and discuss potential systematic effects.  We compute the velocity
dispersion function of clusters in $\S$4 and compare it to the
velocity dispersion function of early-type galaxies in SDSS.  We
discuss our results and conclude in $\S 5$.  We assume $H_0 = 100
h~\kms$, and a flat $\Lambda$CDM cosmology ($\Omega _\Lambda = 1-
\Omega _m$) throughout.  Where not stated explicitly, we assume
$\Omega_m=0.3$ for initial calculations.

\section{The CIRS Cluster Sample}

\subsection{Sloan Digital Sky Survey \label{sdssdesc}}

The Sloan Digital Sky Survey \citep[SDSS,][]{sdss} is a wide-area
photometric and spectroscopic survey at high Galactic latitudes.  The
Fourth Data Release (DR4) of SDSS includes 6670 square degrees of
imaging data and 4783 square degrees of spectroscopic data
\citep{dr4}.

The spectroscopic limit of the main galaxy sample of SDSS is $r$=17.77
after correcting for Galactic extinction \citep{strauss02}.  Assuming
the luminosity function of \citet{blanton03}, the spectroscopic limit
of SDSS corresponds to $M_{^{0.1}r}^*+1$ at $z=0.092$.  The
Cluster And Infall Region Nearby Survey \citep[CAIRNS][]{cairnsi}
found that well-sampled clusters contain dense envelopes known as
caustics \citep{diaferio1999} which provide a direct means of
determining cluster membership.  CAIRNS sampled all galaxies brighter
than about $M^*+1$ and often fainter.  Figure 7 of \citet{cairnsha}
shows the infall patterns for galaxies brighter than $M_{K_s}^*+1$ in
the CAIRNS clusters.  These patterns are readily apparent, but less
well-defined than the full CAIRNS samples \citep{cairnsi}.  Thus, we
expect that infall patterns of clusters with masses similar to CAIRNS
clusters should be apparent to $z\lesssim$0.1, though not much
further.  \citet{cirsi} show the infall patterns of the clusters at
$z\leq$0.1 in SDSS and compute the mass profiles from the caustic
technique and the virial theorem.

Note that the SDSS Main Galaxy Survey is $\sim$85-90\% complete to the
spectroscopic limit.  The survey has $\approx$7\% incompleteness due
to fiber collisions \citep{strauss02}, which are likely more common in
dense cluster fields.  Because the target selected in a fiber
collision is determined randomly, this incompleteness can
theoretically be corrected for in later analysis.  From a comparison
of SDSS with the Millennium Galaxy Catalogue,
\citet{2004MNRAS.349..576C} conclude that there is an additional
incompleteness of $\sim$7\% due to galaxies misclassified as stars or
otherwise missed by the SDSS photometric pipeline.  For our purposes,
the incompleteness is not important provided sufficient numbers of
cluster galaxies do have spectra.

\subsection{X-ray Cluster Surveys \label{xcs}}

Because SDSS surveys primarily low-redshift galaxies, the best sampled
clusters are both nearby and massive.  We therefore search X-ray
cluster catalogs derived from the ROSAT All-Sky Survey for clusters in
DR4.  RASS \citep{rass} is a shallow survey but it is sufficiently
deep to include nearby, massive clusters.  RASS covers virtually the
entire sky and is thus the most complete X-ray cluster survey for
nearby clusters.  

Published cluster catalogs derived from the RASS include the X-ray
Brightest Abell Cluster Survey \citep[XBACS][]{xbacs}, the Bright
Cluster Survey and its extension \citep[BCS and
eBCS][]{bcs,ebcs}, the NOrthern ROSAT All-Sky
galaxy cluster survey \citep[NORAS][]{noras}, and the ROSAT-ESO flux
limited X-ray galaxy cluster survey \citep[REFLEX][]{2001A&A...369..826B}.  We
refer the reader to the individual catalog papers for detailed
descriptions of the construction of the catalogs \citep[][contains a
summary]{cirsi}.  When combined, these catalogs cover virtually the
entire sky at high Galactic latitudes ($|b|>$20$^\circ$) to a flux
limit of $\approx$3$\times 10^{-12}$erg cm$^{-2}$ s$^{-1}$.  None of
these catalogs is complete, so by combining them, we create a
substantially more complete composite catalog.   

\subsection{Definition of the Mass Function Sample}

When multiple X-ray fluxes are available for a cluster, we use the
most recently published value.  The order of preferences is therefore:
REFLEX, NORAS, BCS/eBCS, XBACs.  The details of the flux
determinations in the different catalogs vary somewhat: NORAS and
REFLEX both measure fluxes with Growth Curve Analysis, while BCS/eBCS
uses Voronoi Tessellation and Percolation.  Despite these differences,
the derived fluxes agree fairly well \citep[see Figure 21 of][for a
direct comparison of NORAS and BCS/eBCS]{noras}, so combining the
catalogs is a reasonable procedure.  By using the most recently
published flux value, we use the GCA flux if available and the VTP
flux only when no GCA flux is available.  

We define a flux-limited and
redshift-limited sample of clusters with the criteria $f_X\geq 3\times
10^{-12}$erg s$^{-1}$cm$^{-2}$ (0.1-2.4 keV) and $z\leq$0.10.  Note
that variations in the method of determining flux in different
catalogs may affect the precise flux limit.  The only cluster in XBACS
but not in any other catalog is A1750b: XBACS identifies two X-ray
sources (A1750a/b), whereas other catalogs treat them as one source.
We follow \citet{2004A&A...425..367B} and treat A1750 as a single
source rather than two separate sources as in XBACs.  Similarly, the
galaxy NGC5813 is bound to the NGC5846 group \citep{mahdavi05}.
Because the dynamics of NGC5813 are dominated by the NGC5846 system,
we eliminate NGC5813 from the sample (both are eliminated by our
minimum redshift cutoff below).  We inspect the redshift data around
each cluster to confirm the cluster redshift and find that A2064 has
an incorrect redshift (and X-ray luminosity) listed in NORAS.  The
correct redshift is 0.0738 instead of 0.1076.  We correct the X-ray
luminosity accordingly.  Our final flux and volume limited sample
contains \ncirs ~clusters within the SDSS DR4 spectroscopic footprint.
We will refer to this sample as the CIRS (Cluster Infall Regions in
SDSS) clusters in the rest of this paper.  The completeness of the
CIRS sample is limited by the completeness of the underlying cluster
catalogs.  However, by combining clusters from the various catalogs we
should be more complete than any one of the catalogs.  The local
nature of the sample ($z\leq$0.1) likely improves the completeness of
the CIRS sample relative to a flux-limited sample.  We discuss the
modest potential incompleteness in more detail in
$\S$\ref{systematics}.  The clusters are an unbiased sample: the
selection of the CIRS sample is based purely on X-ray flux and the
footprint of the SDSS DR4 spectroscopic survey.  We confirm this claim
with a $V/V_{max}$ test: we find $<V/V_{max}>$=0.518$\pm$0.035
compared to an expected value of 0.5 for a complete, uniform sample
(see $\S$\ref{systematics}).  A homogeneous, complete, all-sky catalog
of extended X-ray sources in RASS would allow a significant
improvement in the determination of the mass function.  Unfortunately,
no such catalog currently exists.

For computing the mass function, we omit three clusters (Virgo,
NGC4636, and NGC5846) with $z<0.02$ because their peculiar velocities
are likely large compared to their redshifts.  From our analysis of
individual systems in CIRS \citep[see $\S$6 of][]{cirsi}, we
eliminate A1035A/B and A1291A/B as superpositions which would lie
below the flux limit if they were resolved in the catalogs.  Finally,
we eliminate A2249 because its center lies $\approx$2$\arcm$ off the
edge of the SDSS DR4 spectroscopic footprint.  After eliminating these
six clusters, our mass function sample includes 66 clusters.  We test
the effects of returning these clusters to the mass function sample in
$\S$\ref{systematics}.

\section{The Cluster Mass Function}

\subsection{Estimating the Mass Function}

\begin{figure}
\figurenum{1}
\plotone{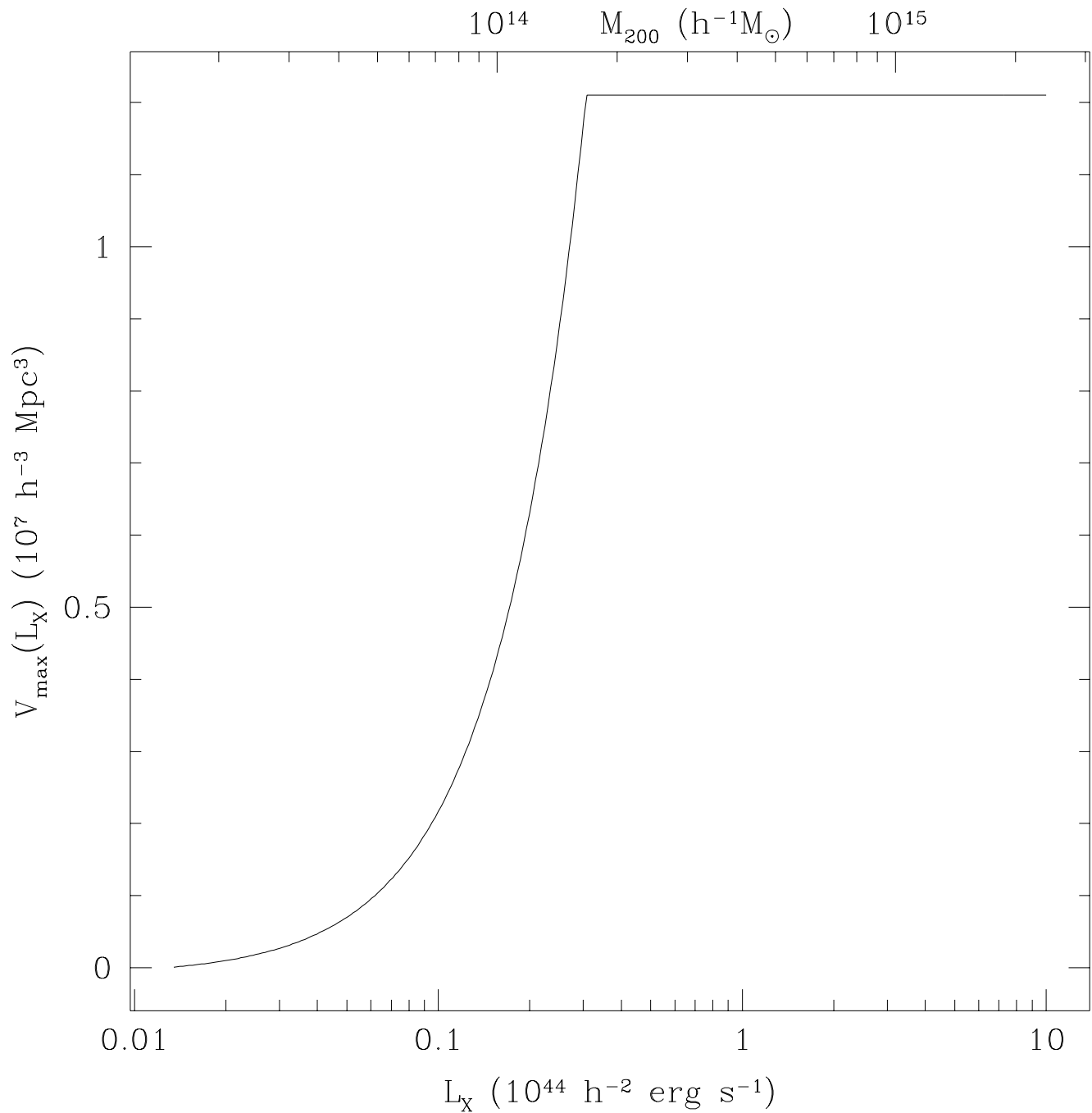}
\caption{\label{vmax} 
The maximum volume sampled by CIRS as a function of X-ray luminosity.  
The top axis shows the equivalent mass assuming the $L_X-M_{200}$ 
relation of the CIRS sample.}
\end{figure}

We compute both the standard cluster mass function
$dn(M)/d\mbox{log}M$ and the cumulative mass function n($>$M), the
number density of clusters more massive than M.  To compute the mass
functions, we use the 1/$V_{max}$ estimator \citep{schmidt68}, where
$V_{max}(L_X)$ is the maximum (comoving) volume a cluster with X-ray
luminosity $L_X$ would lie within our flux and redshift limited
sample.  In each logarithmic mass bin, we sum the clusters
\beqn
\frac{dn(M)}{d\mbox{log}M} = \frac{1}{d\mbox{log}M}\sum_{i} \frac{1}{V_{max}(L_{X,i})}
\eeqn
where the sum is over clusters within the mass bin.  By using
$V_{max}(L_X)$ rather than $V_{max}(M)$, we avoid the need to know the
slope, normalization, and scatter of the scaling relation $L_X-M$ to
properly calculate $V_{max}$ \citep{hiflugcs}.  Figure
\ref{vmax} shows the maximum volume probed by CIRS as a function of
X-ray luminosity.  Although CIRS covers a much smaller area than
HIFLUGCS, $V_{max}(L_X)$ is larger for CIRS for $\mbox{log}L_X<43.8$
because of the smaller flux limit.  We calculate $V_{max}$ assuming a
flat $\Omega_m$=0.3 cosmology.  We repeated our analysis using
$\Omega_m$=0.1 and $\Omega_m$=1 to calculate $V_{max}$ and find that
this has a negligible effect due to the local nature of the sample.
We correct our mass functions for a 5\% incompleteness (see
$\S$\ref{complete}) which we assume applies uniformly to the sample.

\begin{figure}
\figurenum{2}
\plotone{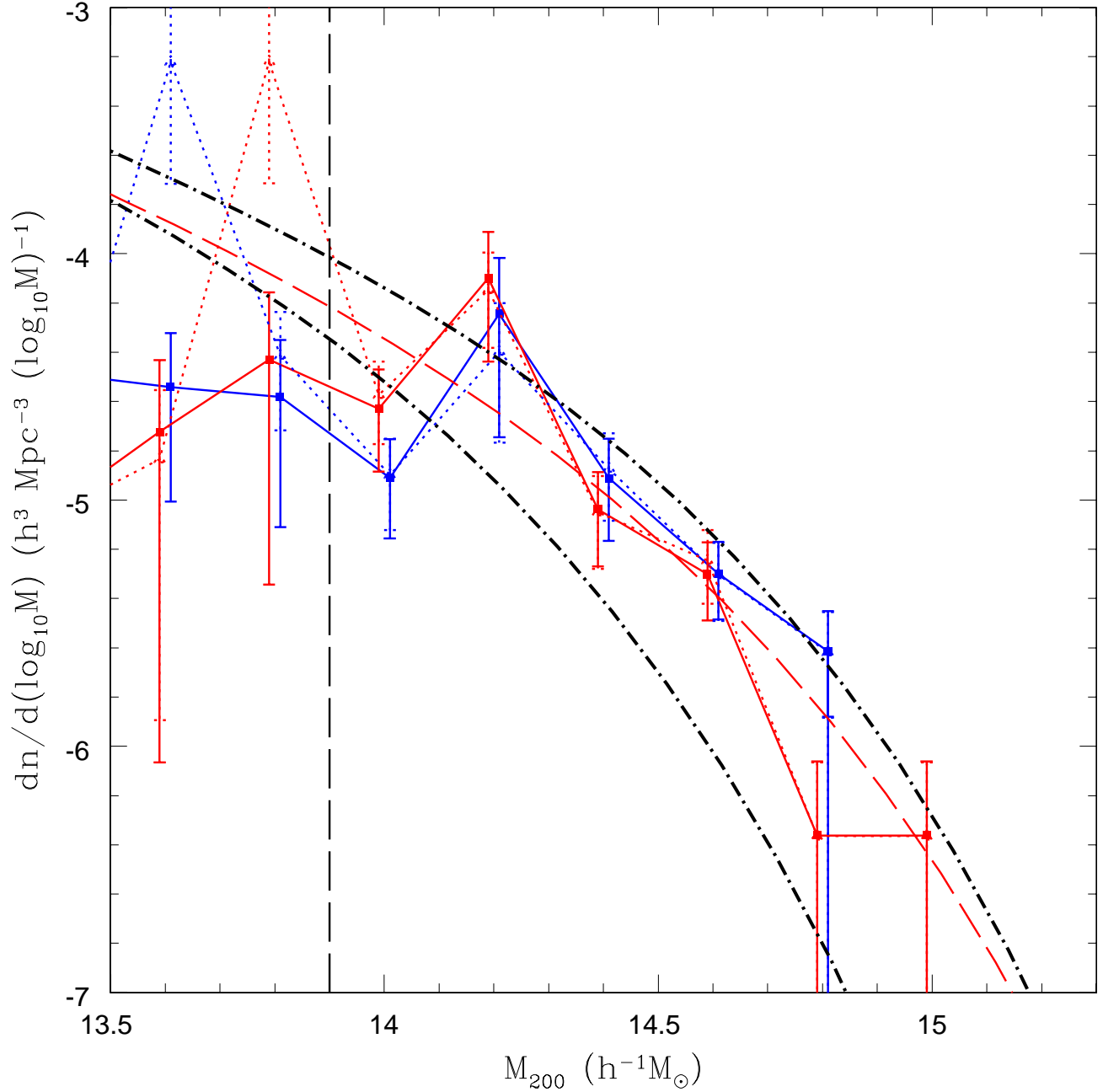}
\caption{\label{dmfn} 
The mass function of the CIRS sample.  Red and blue lines are computed
using virial masses and caustic masses respectively.  The thick
dash-dotted lines show the mass functions computed using the
cosmological parameters from the WMAP1 results (upper) and
WMAP3 results (lower) using the results of
\citet{jenkins01}.  The red dashed line shows the best-fit mass
function for the CIRS virial mass function. The vertical line
indicates the minimum mass we use to constrain cosmological
parameters.  The dotted lines and errorbars show the mass function
computed after removing the minimum redshift and including all
possible mergers as separate systems.  This demonstrates the
importance of cosmic variance at these low masses.}
\end{figure}

We estimate the uncertainty in the mass function as
$[\Delta (dn(M)/d\mbox{log}M)]^2 = \sum_i [1/V_{max}(L_{X,i})^2]$.  Figure \ref{dmfn} shows
the CIRS mass function and Figure \ref{mfn} shows the cumulative mass
function.  We compute the masses in two ways, one using the virial
theorem (red lines in Figures \ref{dmfn} and \ref{mfn}) and the other
(blue lines) using the caustic technique
\citep{diaferio1999,cirsi}.  We use the caustic technique to identify
cluster members in both techniques.  As a membership classification
algorithm, the caustic technique is similar to the ``shifting gapper''
technique \citep[e.g.,][]{1996ApJ...473..670F}, but is less sensitive
to interlopers.  The mass functions computed with these two mass
estimators agree quite well, a confirmation that the caustic technique
provides reliable mass estimates \citep{diaferio1999,gdk99,rines2000,rqcm,rines01a,rines02,bg03,cairnsi,cirsi}.
We estimate the masses $M_{200}$ within $r_{200}$, where $r_\Delta$ is
the radius within which the average density is $\Delta \rho_c$, where
$\rho_c$ is the critical density.  We determine $r_{200}$ directly
from the nonparametric mass profiles.  

For the virial masses, we obtain an initial estimate of $r_{200,vir}$
from the virial mass profiles \citep{cirsi}.  We then apply a
correction factor of 8\% to account for the surface pressure term of
the virial theorem.  This factor is calculated from Equation 8 of
\citet{girardi98} assuming galaxies are on isotropic orbits and an NFW
mass profile \citep{nfw97} with a concentration parameter of
$c=r_{200}/r_s=5$ ($r_s$ is a scale radius in the NFW profile).  The
assumption of isotropic orbits is supported by numerous observations
\citep{2000AJ....119.2038V,cairnsi,lokas03,biviano04,cirsi,benatov06,lokas06}, as is
the value of $c_{200}=5$ \citep{mvfs,lin04,cirsi}.  After this
correction, the enclosed density within $r_{200,vir}$ is now
178$\rho_c$.  We therefore reduce the mass by a further 3.3\% to
obtain our final estimate of $M_{200,vir}$.  The concentration $c$ is
expected to be a weak function of mass, $c\propto M^{-0.13}$
\citep{bullock01}.  This corresponds to a 25-30\% change over the mass
range we probe.  This change has little effect on the mass estimates:
assuming $c$=5 leads to a mass underestimate (overestimate) of 3\%
(5\%) for a cluster with $c=$10 (2.5).  The CIRS clusters are quite
well sampled, with an average of $\approx$50 members per cluster.  The
statistical uncertainties in $M_{vir}$ are on average 13\%.  The
caustic and virial mass functions are in good agreement, further
confirming the good agreement between caustic and virial mass
estimates \citep{cairnsi,diaferio05,cirsi}.

The original prescription for calculating the mass function was
proposed via the Press-Schechter formalism \citep{ps74}.  Numerical
simulations predict relatively more massive systems and fewer less
massive systems than Press-Schechter theory \citep{sheth99}.
\citet{jenkins01} provided fitting formulae for a universal mass
function that can be evaluated for many cosmological models.  In
particular, their mass function accurately reproduces the mass
function of dark matter haloes in the Hubble Volume simulation.
Several investigators conclude that the mass function provided by
Equation (B3) of \citet{jenkins01} is close to a universal mass
function and can thus be used to constrain cosmological parameters
\citep{white02,evrard02b,hu03}.  For this formula, haloes are placed
at the most bound particles, then the radius of the halo is increased
until the enclosed spherical overdensity is 180 times the {\it
background} (not critical) density.  This mass $M_{180b}$ (the ``b''
in the subscript indicates background rather than critical density)
must then be converted to $M_{200}$ assuming an NFW profile with $c$=5
\citep{white02,evrard02b,hu03}.  Note that X-ray mass estimates often
require conversion to $M_{500}$; the conversion to $M_{200}$ is less
of an extrapolation and less sensitive to the assumed concentration
parameter.  

Figure \ref{dmfn} shows the mass functions
for the best-fit cosmological parameters from WMAP1
\citep{spergel03} and WMAP3 \citep{spergel06} at
$z$=0.062 (the mean redshift of our sample).  The WMAP mass functions
in Figure \ref{dmfn} are convolved with our mean mass uncertainty
$\sigma_{\mbox{log}M}=0.056$ for appropriate comparison with the
observed mass functions.  Some of the largest differences between the
WMAP1 and WMAP3 were in $\sigma_8$ and
$\Omega_m$, the two parameters which have the largest impacts on the
mass function.  Figure \ref{dmfn} shows that the CIRS mass function
lies between the WMAP1 and WMAP3 results.
\citet{reiprich06} showed that the HIFLUGCS mass function estimated
using the X-ray properties of clusters lies closer to the 
WMAP3 results \citep{hiflugcs}, although other mass functions derived
with X-ray mass tracers find values closer to the WMAP1
results \citep{pierpaoli03,viana03}.

\begin{figure}
\figurenum{3}
\epsscale{0.8}
\plotone{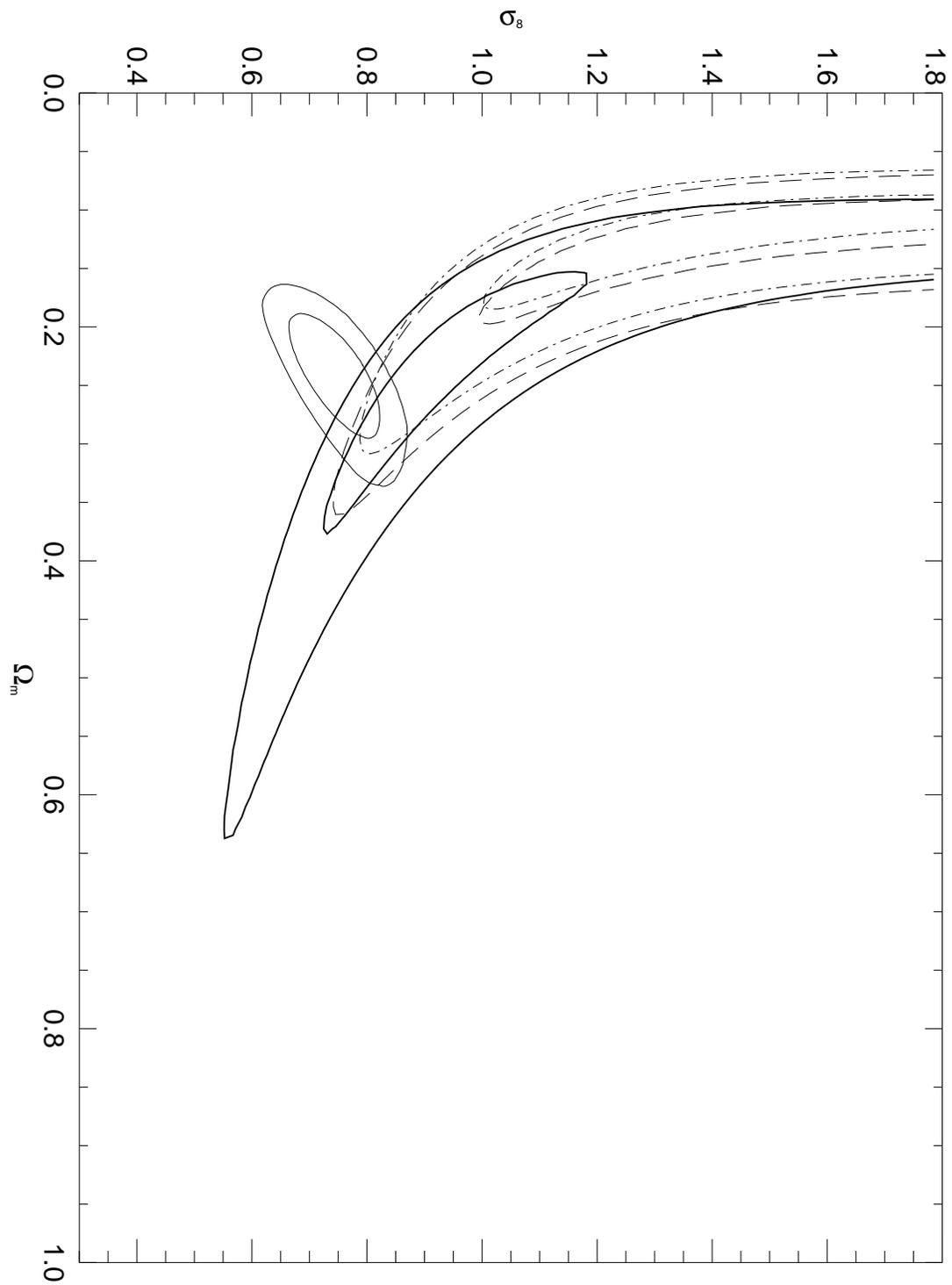}
\caption{\label{omsigsysmass} 
Cosmological constraints from the CIRS mass function. Solid contours
show 1 and 3$\sigma$ confidence levels for $\Omega_m$ and $\sigma_8$
from the virial mass function.  Dashed contours show the constraints
from the caustic mass function.  Dash-dotted contours show the
constraints from the virial mass function with virial masses computed
with only red galaxies.  The solid contours extending to the lower 
left show the 68\% and 95\% confidence levels from WMAP3.  }
\end{figure}

\begin{figure}
\figurenum{4}
\epsscale{0.8}
\plotone{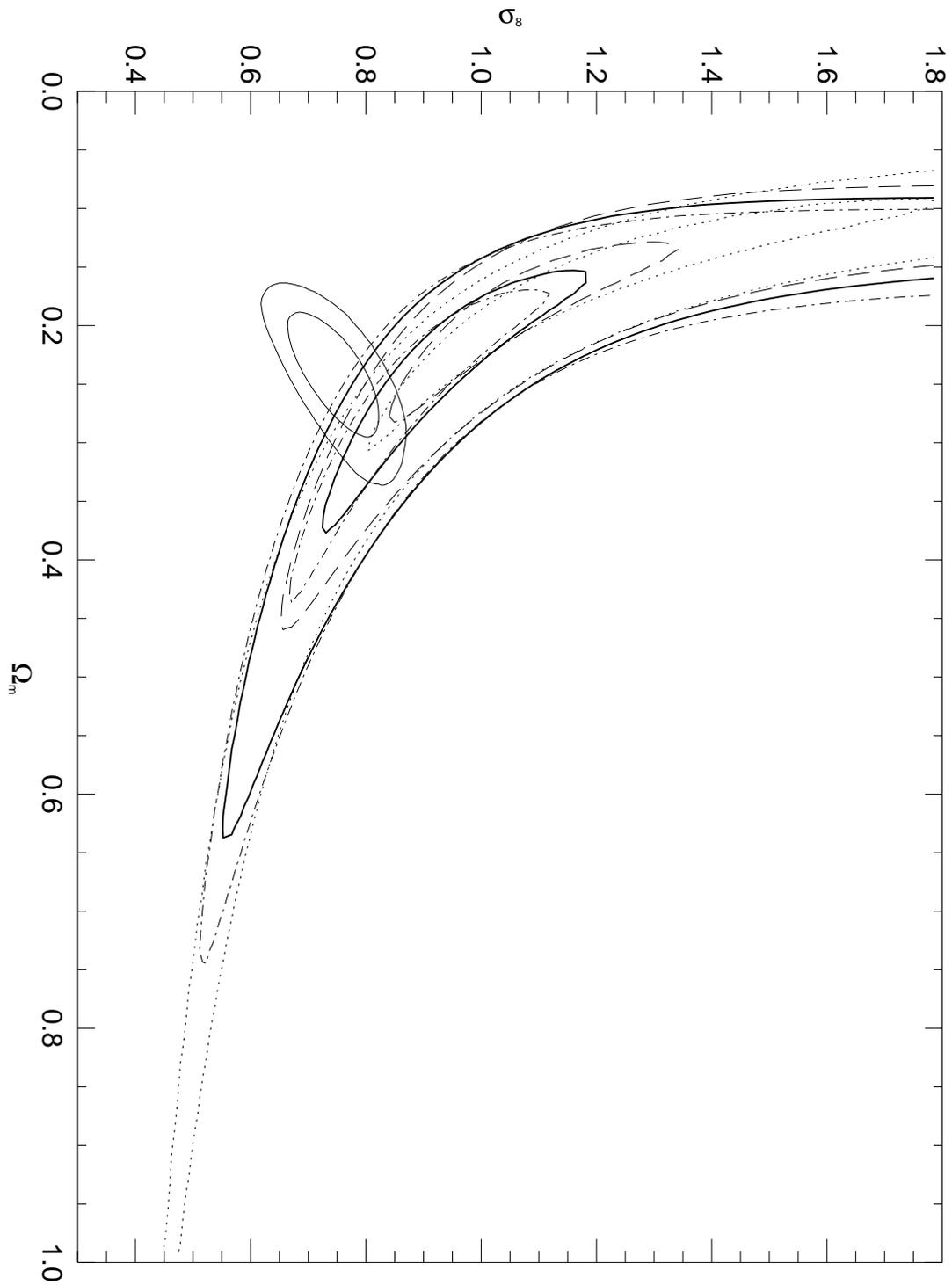}
\caption{\label{omsigh} 
Dependence of cosmological constraints from the CIRS mass function on
the assumed value of $H_0$. Solid contours show 1-3$\sigma$ confidence
levels for $\Omega_m$ and $\sigma_8$ for $h=0.7$.  Dashed and
dash-dotted contours show the 1-3$\sigma$ confidence levels for $h=1$
and $h=0.5$ respectively.  Dotted contours show the 1-3$\sigma$ contours assuming $h$=0.7 and a fixed value of $\Gamma$=0.21.   The solid contours extending to the lower 
left show the 68\% and 95\% confidence levels from WMAP3.}
\end{figure}

\begin{figure}
\figurenum{5}
\plotone{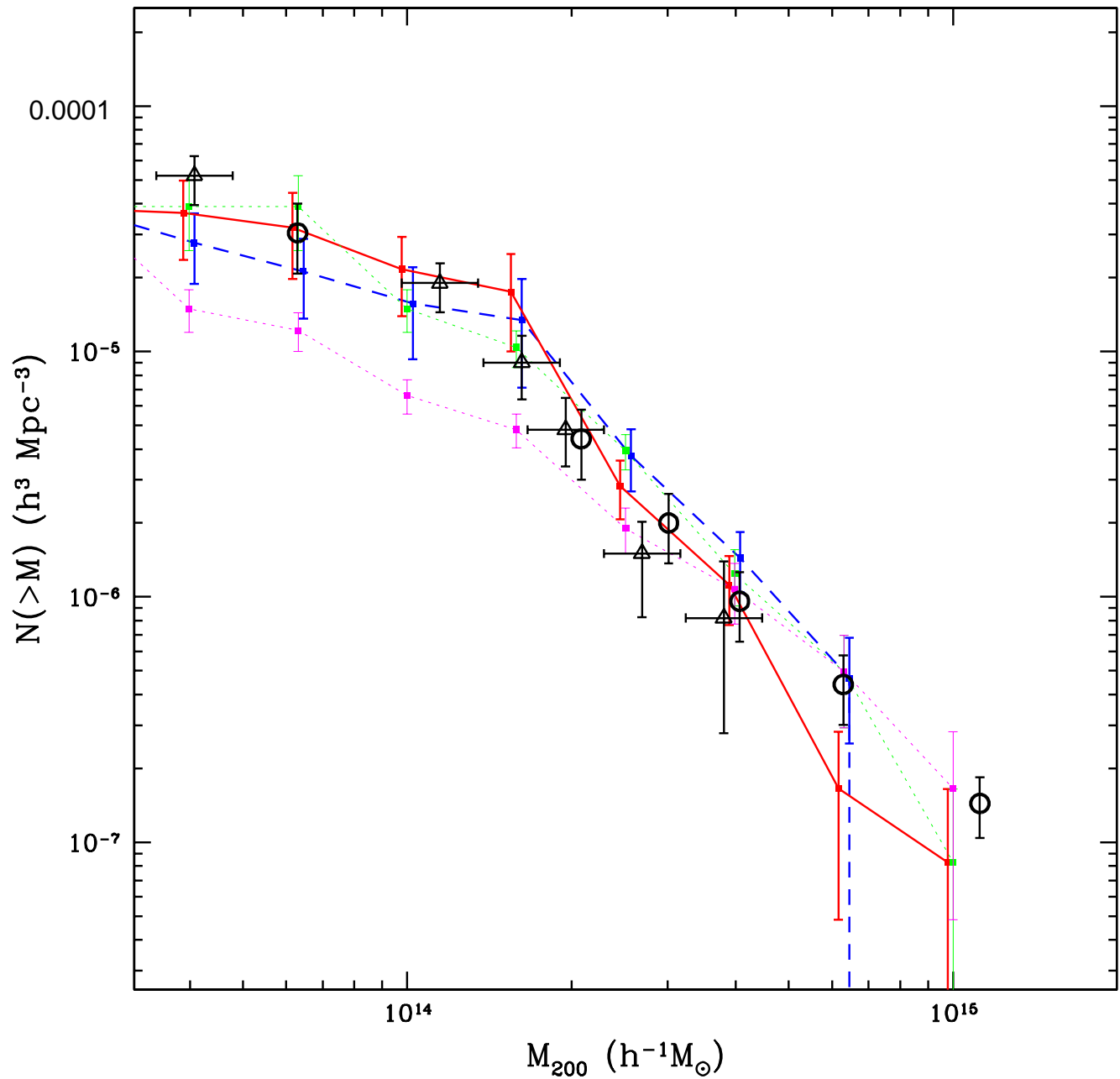}
\caption{\label{mfn} 
Cumulative mass function of the CIRS sample.  Red and blue lines are
computed using virial masses and caustic masses respectively.  Open
circles show the HIFLUGCS mass function computed with X-ray masses
\citep{hiflugcs} and triangles show the mass function computed from early
SDSS data using a mass-richness relation \citep{bahcall03a}.  Green
and magenta lines show the mass function of the CIRS sample computed
with the $L_X-M_X$ relation of \citet{popesso04} and the $L_X-M_{vir}$
relation of \citet{cirsi} respectively. }
\end{figure}

The dotted lines in Figure \ref{dmfn} show the mass function computed
without imposing a minimum redshift on the sample.  Clearly, the
systematic uncertainty in the mass function due to cosmic variance
becomes large at M$\lesssim$10$^{14}\msun$ (see also Figure
\ref{vmax}).  We minimize $\chi^2$ for the CIRS mass function in the
mass range $\mbox{log}M_{200}$=[13.9,15.1] by calculating the
\citet{jenkins01} mass function for a given pair of
($\Omega_m$,$\sigma_8$) and shifting the mass scale from $M_{180b}$ to
$M_{200}$.  We set $\Gamma$ according to
\beqn
\label{gamma}
\Gamma(\Omega_m,h) = \Omega_m h \left(\frac{2.7 K}{T_0}\right)^2 \exp{(-\Omega_b-\sqrt{2 h} \frac{\Omega_b}{\Omega_m})}
\eeqn
\citep{sugiyama95} with $T_0=2.726 K$, $h=0.7$ and $\Omega_b=0.0223 h^2$ 
\citep{spergel06}.  Holding $\Gamma$=0.21 fixed does not strongly affect 
our results.  We convolve the mass function with our mean mass
uncertainty $\sigma_{\mbox{log}M}=0.056$ according to
\begin{eqnarray}
\label{vmaxweight}
d\tilde{n}(M)/dM = \frac{1}{V_{max}(M)}\int_{-\infty}^{+\infty} \frac{dn(M')}{dM'}V_{max}(M') 
\nonumber \\ 
\,\,\,\,\,\,\,\,\,\,\,\, \times (2\pi\sigma_{log M}^2)^{-1/2} \mbox{exp}\left[ \frac{-(\mbox{log}M'-\mbox{log}M)^2}{2\sigma_{log M}^2} \right] d \mbox{log} M'
\end{eqnarray}
where $V_{max}(M)$ is computed using the scaling relation 
\beqn
\mbox{log}(M_{200}h/M_\odot) = 0.763 \mbox{log}(L_{X,44}) + 14.62
\eeqn
from \citet{cirsi} where $L_{X,44}$ is the X-ray luminosity in units
of $10^{44}h^{-2}\mbox{erg}\mbox{s}^{-1}$.  This relation has a
different normalization than the relation found by \citet{popesso05},
but is consistent with the relation of \citet{hiflugcs}.  The
weighting by $V_{max}$ is required because the CIRS mass function is
flux-limited for $z<0.1$.  We repeat the fits without weighting by
$V_{max}$ and find negligible differences.  At smaller masses, the
volume probed is fairly small (Figure \ref{vmax}), so the mass
function estimate is subject to significant uncertainty due to cosmic
variance \citep{hu03}.  In addition, the number of clusters at these
low masses is probably not large enough to sample the scatter in
$L_X$.  Figure \ref{omsigsysmass} shows the 68\% and 99.7\% confidence
levels for $\Omega_m$ and $\sigma_8$ inferred from the virial mass
function (solid contours) and the caustic mass function (dashed
contours).  With the CIRS virial mass function, we estimate
$\Omega_m=0.24^{+0.14}_{-0.09}$ and $\sigma_8=0.92^{+0.24}_{-0.19}$,
or $\sigma_8=0.84\pm$0.03 when holding $\Omega_m=0.3$ fixed (all
assuming $h$=0.7).  The open squares in Figure \ref{omsigsysmass} show
the WMAP1 and WMAP3 estimates with their 68\% uncertainties as
errorbars \citep[][Table 2]{spergel06}.  The CIRS mass function is
consistent with the WMAP3 results within their 95\% confidence levels,
while the WMAP1 results lie inside the 68\% contour of the CIRS mass
function.  To compute the constraints in Figure \ref{omsigsysmass} we
assumed $h$=0.7 and computed $\Gamma$ from Equation (\ref{gamma}).
Figure \ref{omsigh} shows the effects of assuming different values of
$h$ or of assuming a fixed value of $\Gamma$=0.21. Clearly, the mass
function computed with virial masses is now able to provide
cosmological constraints competitive with those from X-ray mass
estimates and even WMAP.

\subsection{Comparison With Other Methods}

We now compare the CIRS mass function and the resulting constraints on
cosmological parameters with previous studies.

\subsubsection{Mass Function}

Figure \ref{mfn} shows the CIRS cluster mass function compared to
other methods.  HIFLUGCS uses the temperature and deprojected
distribution of X-ray gas to estimate cluster masses from the ROSAT
All-Sky Survey \citep{hiflugcs}.  \citet{bahcall03a} use a
mass-richness relation to estimate the masses of clusters from the
Early Data Release of SDSS.  The CIRS mass function agrees well with
both of these independent methods, indicating that the cluster mass
function is robustly estimated at low redshift.

\subsubsection{Cosmological Constraints \label{cosmocompare}}

\begin{figure}
\figurenum{6}
\epsscale{0.7}
\plotone{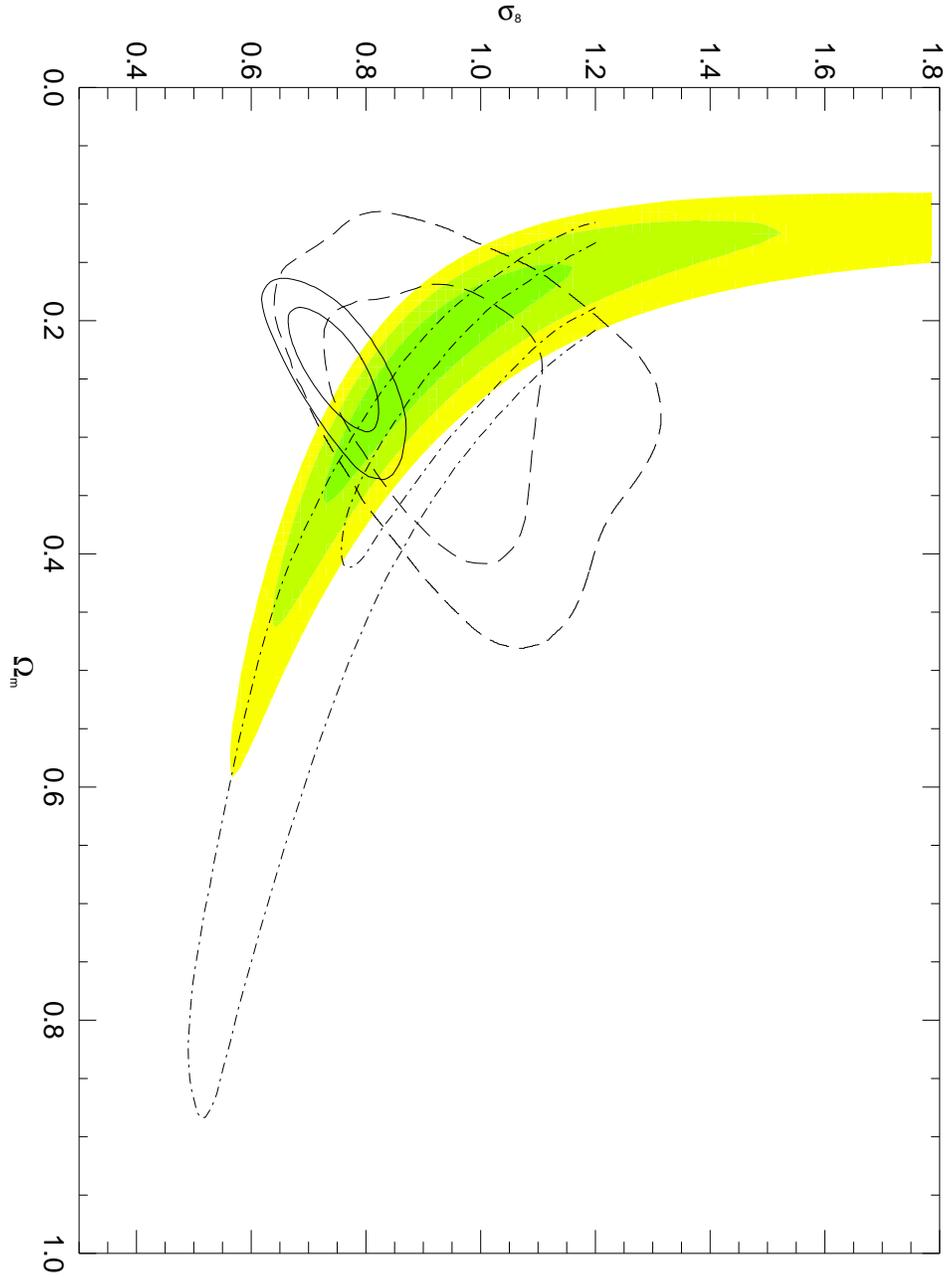}
\caption{\label{omsig} 
Cosmological constraints from the CIRS virial mass function compared
to other results. Colored contours show 1-2-3$\sigma$ confidence
levels for $\Omega_m$ and $\sigma_8$.  The dashed, solid, and dash-dotted contours show
the 68\% and 95\% confidence levels from WMAP1, WMAP3, and the CFHTLS
Wide survey respectively \citet[adapted from Figures 1 and 7 of][]{spergel06}.  }
\end{figure}

\begin{figure}
\figurenum{7}
\epsscale{0.7}
\plotone{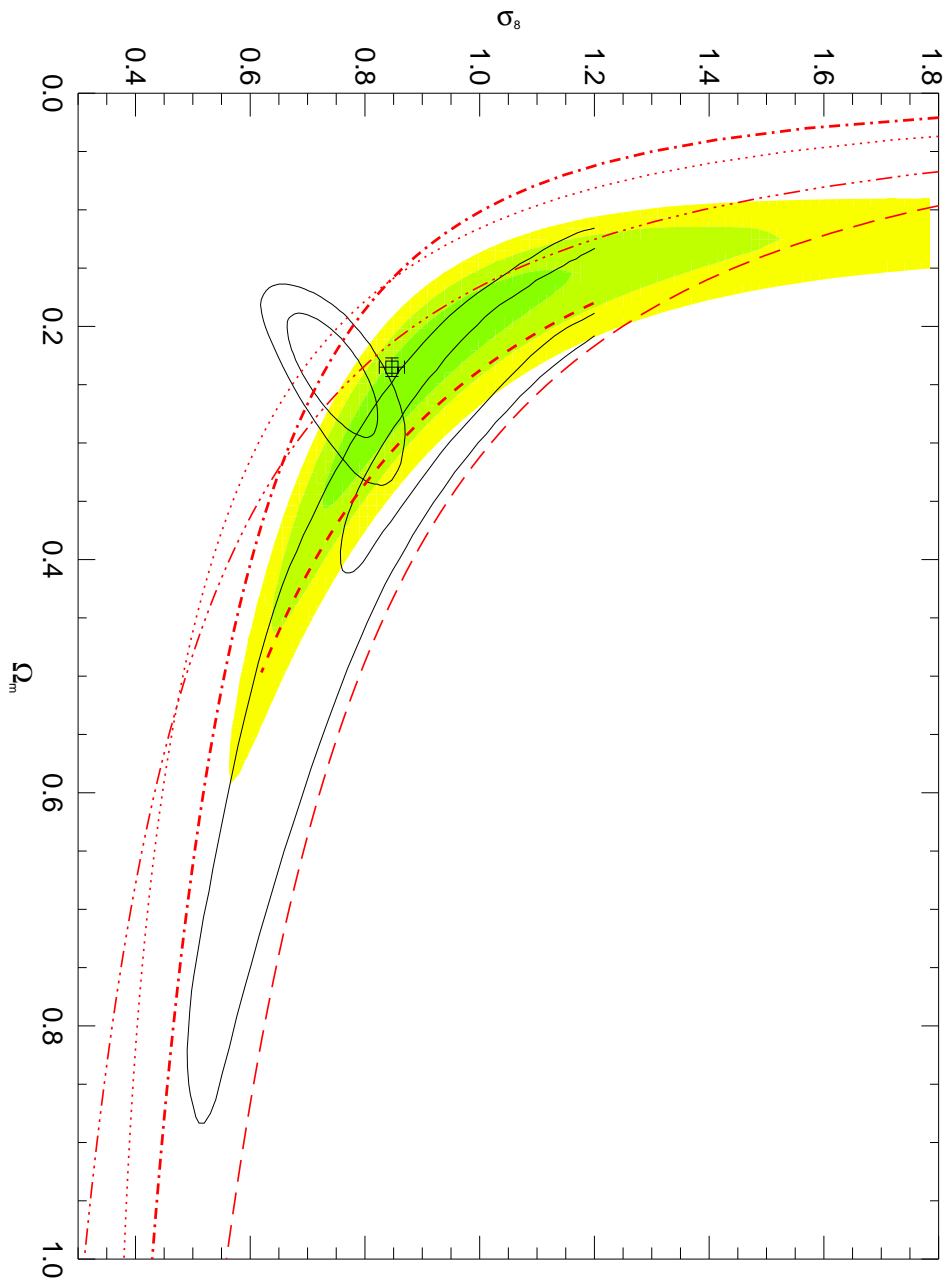}
\caption{\label{omsignew2} 
Cosmological constraints from the CIRS virial mass function compared
to other results. Colored contours show 1-2-3$\sigma$ confidence
levels for $\Omega_m$ and $\sigma_8$.  Dash-triple-dotted \citep{heymans05} and long-dashed \citep{massey05} lines show the range of constraints from
cosmic shear measurements. The thick red short-dashed line shows the cosmic
shear estimate from the CFHT Legacy Survey ``Deep'' sample
\citep{semboloni06}.    The red dotted line shows the 
constraints from the mass function inferred from the $M-L_X$
relation obtained from weak lensing measurements in SDSS
\citep{viana02}. The thick red dash-dotted line shows the
constraint of \citet{dahle06} using weak lensing mass estimates
of X-ray selected galaxy clusters.   The open square with extremely small errorbars is 
the result of \citet{seljak06} for joint constraints from WMAP3,
small-scale CMB measurements, SDSS galaxy clustering, supernovae, and
the Lyman-$\alpha$ forest.   The solid contours extending to the lower 
left show the 68\% and 95\% confidence levels from WMAP3.} 
\end{figure}

\begin{figure}
\figurenum{8}
\epsscale{0.7}
\plotone{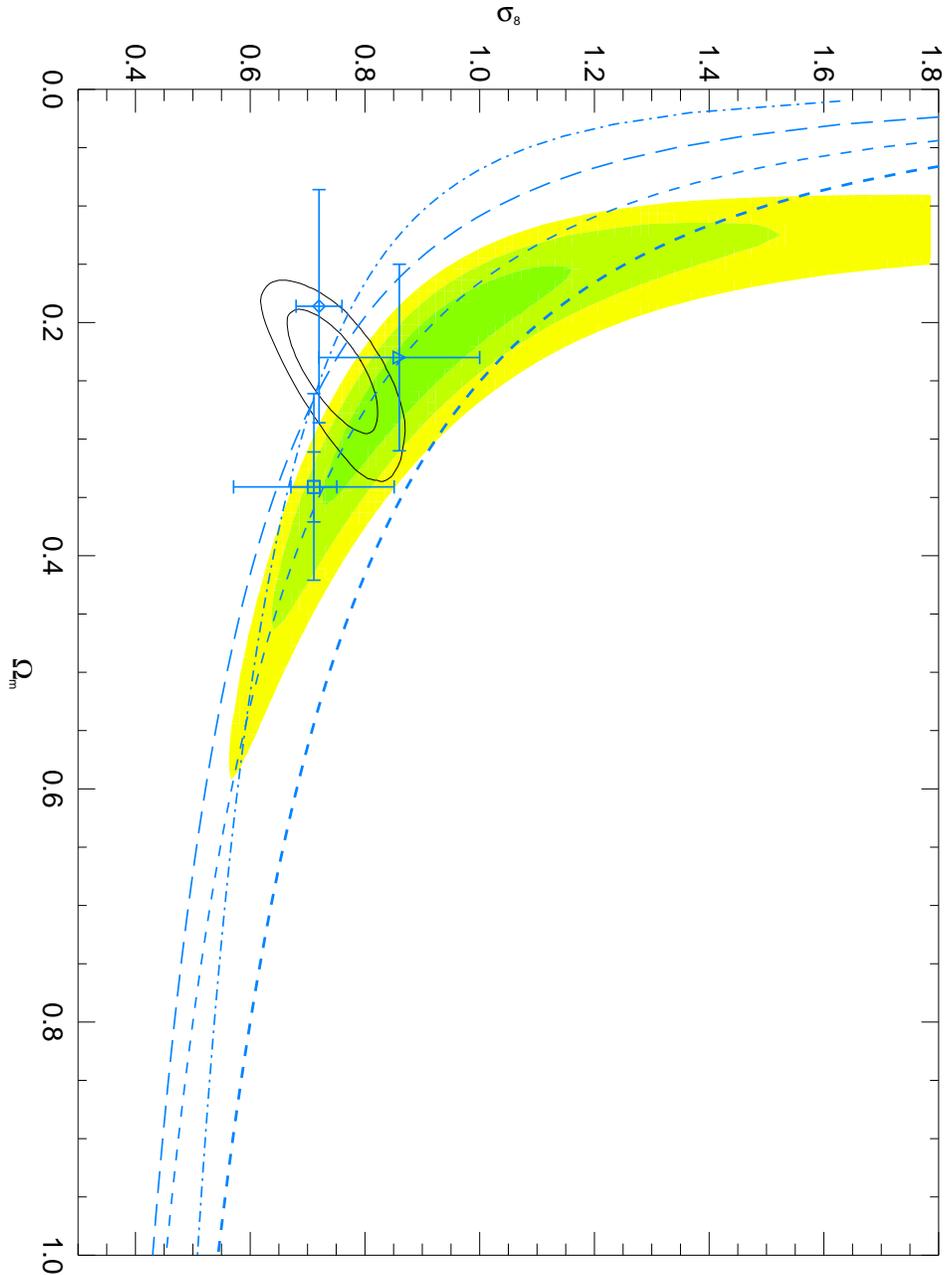}
\caption{\label{omsignew3} 
Cosmological constraints from the CIRS virial mass function compared
to other results. Colored contours show 1-2-3$\sigma$ confidence
levels for $\Omega_m$ and $\sigma_8$. 
Blue lines show constraints from X-ray
cluster data.  The lower short-dashed line is the temperature function
constraint of \citet{seljak02}; the upper short-dashed line shows
this relation increased by 20\% in $\sigma_8$ to show the systematic
offset suggested by \citet{rasia05}.  The long-dashed line is from the
X-ray mass function of \citet{hiflugcs}, and the dash-dotted line
shows the constraint from the X-ray luminosity function
\citep{allen03}.  The diamond shows a recent X-ray measurements
from \citet[][]{vv04} using the baryon mass function and baryon
fraction.  The triangle shows the constraint from
\citet[][]{pierpaoli03} using the X-ray temperature function. The
square at large $\Omega_m$ is from combining the cluster abundance
with the observed clustering \citep[][the two sets of errorbars show
statistical and systematic errors]{schuecker03}.  All error bars
denote 68\% uncertainties.   The solid contours extending to the lower 
left show the 68\% and 95\% confidence levels from WMAP3.} 
\end{figure}

\begin{figure}
\figurenum{9}
\epsscale{0.7}
\plotone{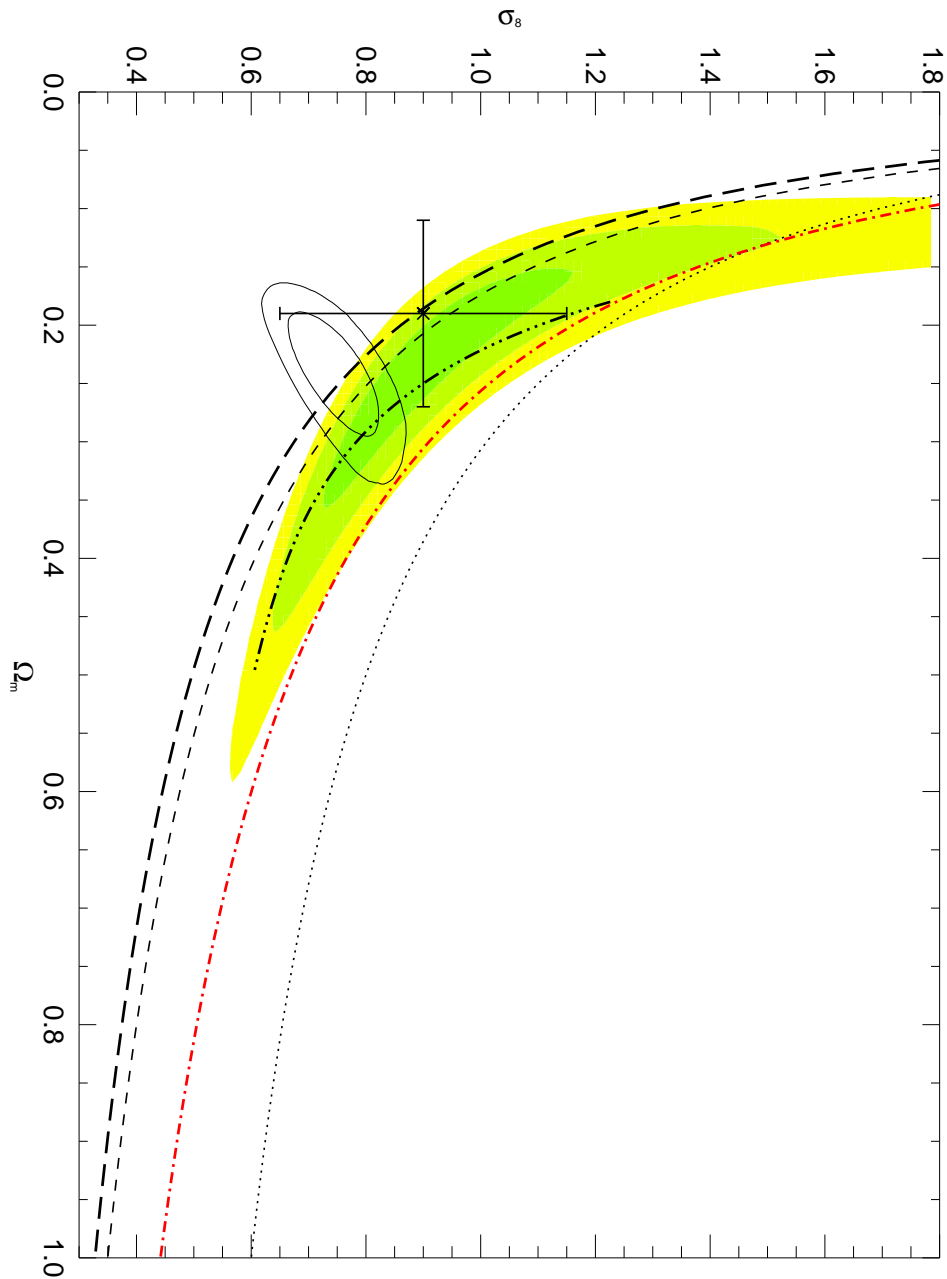}
\caption{\label{omsignew4} 
Cosmological constraints from the CIRS virial mass function compared
to other results. Colored contours show 1-2-3$\sigma$ confidence
levels for $\Omega_m$ and $\sigma_8$.  
Black lines show constraints from
optical cluster data.  The upper dotted black line  is the virial mass
function of \citet{girardi98}, and the thick black short-dashed line
(and black cross) is from the mass-richness relation
\citep{bahcall03b}.  The triple-dot-dashed line is the
constraint from \citet{eke05} for galaxy groups in the 2dFGRS.  The
long-dashed line shows the constraints of \citet{tinker05} from
modeling the halo occupation distribution of galaxies and cluster
mass-to-light ratios.  The red dash-dotted line shows the
constraint from combining peculiar velocity surveys \citep{pike05}.
All error bars denote
68\% uncertainties.   The solid contours extending to the lower 
left show the 68\% and 95\% confidence levels from WMAP3. }
\end{figure}

Figures \ref{omsig}-\ref{omsignew4} show a variety of other
constraints in the $(\Omega_m,\sigma_8)$ plane.  The colored contours
in each Figure are for the CIRS virial mass function with the 68\%,
95\%, and 99.7\% confidence levels for $\Omega_m$ and $\sigma_8$.  The
two sets of contours shown in each panel show the 68\% and 95\%
confidence levels from WMAP3.  The dashed contours in Figure
\ref{omsig} shows these levels for WMAP1, and the dash-dotted contours
are the confidence levels from CFHTLS \citep{hoekstra06}.  Note that
the contours for WMAP1 are adapted from Figure 1 of \citet{spergel06},
which displays the contours in the $\Omega_m h^2$-$\sigma_8$ plane.
We increase the uncertainties in $\Omega_m$ to account for the
uncertainty in $h$.  The 68\% confidence levels of WMAP3 and CIRS
overlap each other, demonstrating excellent agreement between these
two very different types of constraints.  The CIRS contours lie closer
to the WMAP3 contours than the CFHTLS results.

Figure \ref{omsignew2} compares the CIRS results with various lensing
constraints.  Red lines show the range of cosmic shear measurements,
from the ``low'' result of \citet[][dash-triple-dotted
line]{heymans05} to the ``high'' result of \citet[][long-dashed
line]{massey05}.  The open square with extremely small errorbars is
the result of \citet{seljak06} for joint constraints from WMAP3,
small-scale CMB measurements, SDSS galaxy clustering, supernovae, and
the Lyman-$\alpha$ forest.  The 68\% confidence contour of the CIRS
virial mass function encloses this point.  Remarkably, the cosmic
shear estimate from the CFHT Legacy Survey ``Deep'' sample
\citep[][thick red short-dashed line in Figure
\ref{omsignew2}]{semboloni06} lies within the CIRS mass function 68\%
contour, and the CFHTLS ``Wide'' sample overlaps significantly with
the CIRS contours.  The red dotted line shows the 
constraints from the mass function inferred from the $M-L_X$
relation obtained from weak lensing measurements in SDSS
\citep{viana02}.  Finally, the thick red dash-dotted line shows the
recent constraint of \citet{dahle06} using weak lensing mass estimates
of X-ray selected galaxy clusters.  These constraints lie somewhat
below the CIRS constraints and closer to the WMAP3.  One potential
difference between the samples is that the lensing clusters are at
$z$=0.15-0.3, so cluster evolution may cause some of this difference.

Blue lines in Figure \ref{omsignew3} show constraints from X-ray
measurements.  The lower short-dashed line is the temperature function
constraint of \citet{seljak02}; the upper short-dashed line shows
this relation increased by 20\% in $\sigma_8$ to show the systematic
offset suggested by \citet{rasia05}.  The long-dashed line is from the
X-ray mass function of \citet{hiflugcs}, and the dash-dotted line
shows the constraint from the X-ray luminosity function
\citep{allen03}.  The diamond shows a recent X-ray measurements
from \citet[][]{vv04} using the baryon mass function and baryon
fraction.  The triangle shows the constraint from
\citet[][]{pierpaoli03} using the X-ray temperature function. The
square at large $\Omega_m$ is from combining the cluster abundance
with the observed clustering \citep[][the two sets of errorbars show
statistical and systematic errors]{schuecker03}.  All error bars
denote 68\% uncertainties.

The upper dotted black line in Figure \ref{omsignew4} is the virial mass
function of \citet{girardi98}, and the thick black short-dashed line
(and black cross) is from the mass-richness relation
\citep{bahcall03b}.  The thick triple-dot-dashed line is the
constraint from \citet{eke05} for galaxy groups in the 2dFGRS.  This
line passes close to the middle of the CIRS 68\% contour.  The black
long-dashed line shows the constraints of \citet{tinker05} from
modeling the halo occupation distribution of galaxies and cluster
mass-to-light ratios.  Finally, the dash-dotted line shows the
constraint from combining peculiar velocity surveys \citep{pike05}.

The constraints on ($\Omega_m$,$\sigma_8$) from the CIRS virial mass
function are in excellent agreement with previous results.  They lie
between the upper and lower limits of cosmic shear estimates, and very
close to the CFHTLS constraints.  Compared to optical estimates, CIRS
lies slightly above \citet{bahcall03b}, but agrees extremely well with
\citet{eke05}, who used a simplified virial-like equation to estimate
masses.  The CIRS constraints yield slightly larger values of
($\Omega_m$,$\sigma_8$) than most X-ray estimates, but the systematic
temperature underestimate suggested by \citet{rasia05} would bring
these results into reasonable agreement with the CIRS constraints.
Finally, the constraint of \citet{seljak06} combines WMAP3 with a
number of independent probes to obtain estimates of $\Omega_m$ and
$\sigma_8$ with very small statistical uncertainties.  The values of
these parameters are shifted noticeably relative to the constraints
from WMAP3 alone, and the sense of this shift is exactly in the
direction of yielding better agreement with the CIRS mass function.

\subsection{Testing for Systematics \label{systematics}}

It is tempting to conclude from the above comparisons that the CIRS
virial mass function is an unbiased, reliable estimator of $\Omega_m$
and $\sigma_8$.  Below, we overcome this temptation and discuss 
potential systematic effects which could bias our results.

\subsubsection{Completeness of CIRS \label{complete}}

The CIRS sample is derived from four independent X-ray cluster
catalogs constructed from the ROSAT All-Sky Survey.  None of these
surveys is complete: BCS/eBCS is estimated to be 80-90\% complete, the
main NORAS catalog is estimated to be 52\% complete, the NORAS
extended sample in the $9^h$-$14^h$ ($\delta\geq$0) region is 82\%
complete (both NORAS estimates are obtained by comparing source counts
to the deeper REFLEX survey and assuming identical log N-log S
distributions in the northern and southern hemispheres).  XBACS
searched for extended X-ray emission around all Abell clusters, so the
CIRS sample should include all Abell and ACO clusters \citep{aco1989}
with significant X-ray emission (although large offsets between Abell
and X-ray centers may cause XBACS to miss some systems).  To estimate
the approximate incompleteness in CIRS, we note that 53 of the 66 CIRS
clusters are in the northern hemisphere ($\delta\geq$0) and thus in
the area of sky covered by NORAS.  Of these 53 clusters, 29 are NORAS
(main catalog) clusters meeting our flux limit.  If we assume that the
52\% completeness of the main NORAS catalog applies to our
redshift-limited subsample, there should be (29/0.52)=56 northern
clusters in CIRS.  Thus, the completeness of the composite CIRS
catalog is $\approx$95\%.

We refit the mass function using only clusters in the REFLEX and NORAS
main catalogs (omitting those in the NORAS 9$^h$-14$^h$ sample).  The
REFLEX southern catalog covers a small fraction of the DR4 area;
including it eliminates the need to recalculate the effective area of
the survey.  The REFLEX/NORAS sample should have a completeness
similar to, but slightly larger than, the 52\% completeness of the
NORAS main sample.  The cosmological constraints from this subsample
assuming either 52\% or 60\% completeness are very similar to the
constraints from the entire sample, although the confidence contours
are larger.  This result suggests that the use of the composite
catalog of X-ray clusters introduces no significant bias in the mass
function.

\citet{popesso04} searched several X-ray catalogs of clusters and 
groups including the unpublished NORAS2 catalog, a deeper and more
complete version of NORAS.  They find four additional systems in the
SDSS DR3 spectroscopic area which are not in any of the X-ray cluster
catalogs used to construct the CIRS sample.  Of these four systems,
one (A1139) is included in the NORAS $9^h$-$14^h$ catalog but lies below our
flux limit, another (NGC6173=A2197E) is sometimes included as part of
A2197W, and \citet{rines02} find a flux for A2197E that is below the
CIRS flux limit.  The remaining two systems are apparently X-ray
groups with velocity dispersions of $\sigma_p$$\lesssim$450$\kms$ and
virial masses $\lesssim$1$\times$10$^{14} \msun$.  This confirms that
many of the systems missing from the CIRS sample are of relatively low
mass and would probably lie below our adopted mass cutoff.

We apply the $V/V_{max}$ test \citep{schmidt68} to the CIRS mass
function sample to test for incompleteness.  We find
$<$$V/V_{max}$$>$=0.518$\pm$0.035, compared to the expected value of 0.5
for a uniform sample.  This test shows that the CIRS sample is a
representative sample.

\begin{figure}
\figurenum{10}
\epsscale{0.8}
\plotone{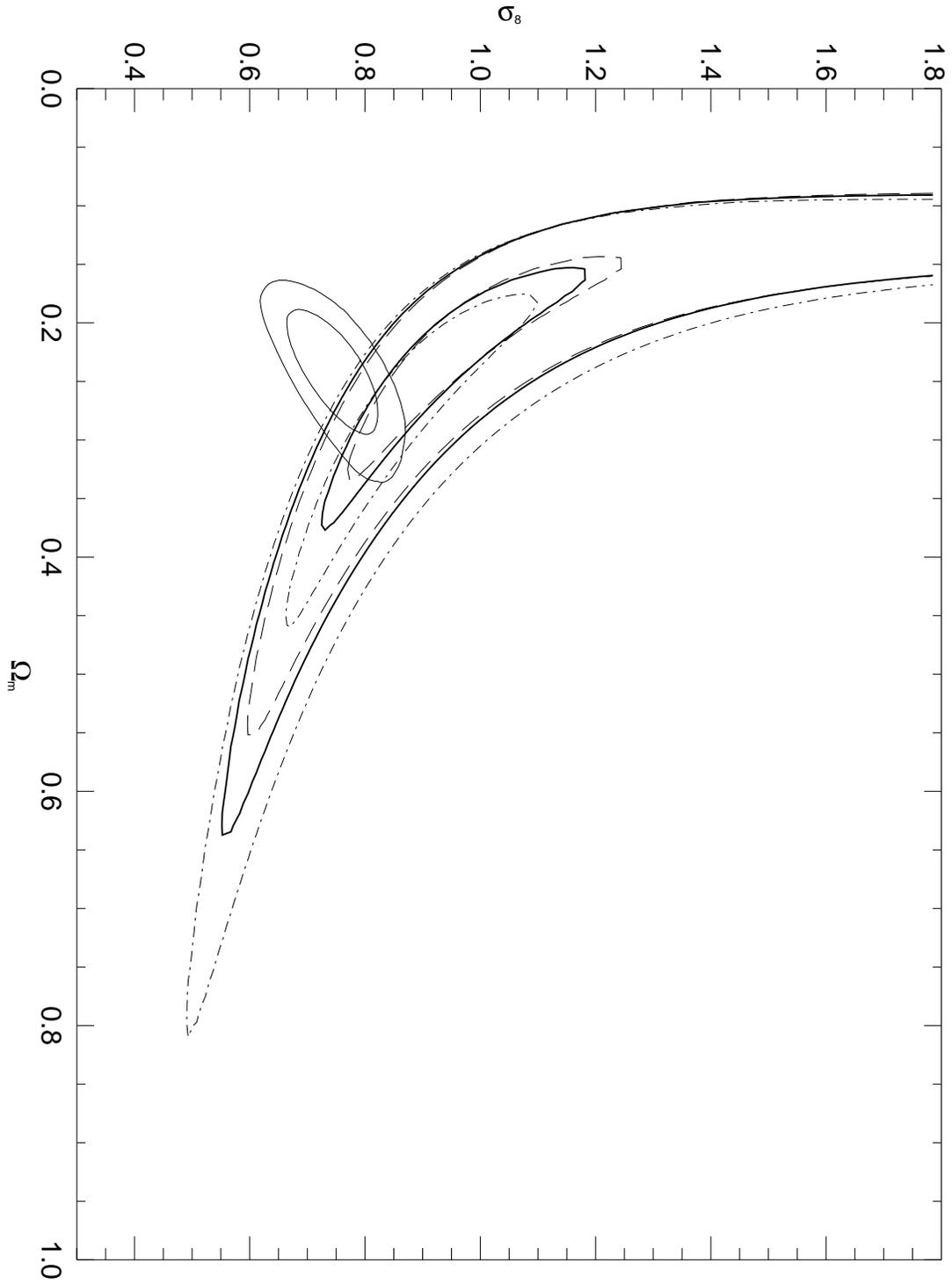}
\caption{\label{omsigsyscuts} 
Cosmological constraints from the CIRS virial mass function.  Solid
contours show 1-3$\sigma$ confidence levels for $\Omega_m$ and
$\sigma_8$.  Dash-dotted and dashed contours show the 1-3$\sigma$
confidence levels from adopting a higher flux limit and from including
both cluster ``pairs'' and low-redshift systems respectively.   
The solid contours extending to the lower 
left show the 68\% and 95\% confidence levels from WMAP3.} 
\end{figure}

We recalculate the cluster mass function using a higher flux limit
than the fiducial value of $3\times 10^{-12}$erg s$^{-1}$cm$^{-2}$
(0.1-2.4 keV) cited for the parent X-ray catalogs.  The mass function
computed from the 46 clusters brighter than $f_X\geq 5\times
10^{-12}$erg s$^{-1}$cm$^{-2}$ (0.1-2.4 keV) is almost identical to
our standard mass function. Figure \ref{omsigsyscuts} shows the effect
of adopting this higher flux limit on our cosmological constraints
(dash-dotted contours).  The confidence contours get larger, but the
best-fit parameters do not change significantly.



The minimum redshift we imposed to select the CIRS mass function
sample reduces the number of low-mass systems sampled.  Including
these systems causes a substantial increase in the number of systems
at masses $\lesssim$1$\times$10$^{14} \msun$, confirming our assertion
that cosmic variance becomes significant in this mass range for the
CIRS sample.

Distinguishing a single, unrelaxed cluster from two merging clusters
is a difficult observational problem.  The CIRS sample contains
several examples of multiple clusters \citep{cirsi}.  We test the
sensitivity of the CIRS mass function to this effect by making the
extreme assumptions that either none or all multiple peaks are
separate halos.  The affected systems are A1035A, A1035B, A1291A,
A1291B, A1750A, A1750B, A2197W, A2197E, NGC5813, and NGC4636.  Figure
\ref{omsigsyscuts} shows that including all of these systems as well
as the low-redshift systems (we call this the ``maximal'' CIRS mass
function) makes little difference to our constraints in the
($\Omega_m$,$\sigma_8$) plane.

A related completeness problem stems from the fact that, although we
use galaxy dynamics to estimate masses, the CIRS sample is selected by
X-ray flux (with an upper limit on redshift).  This selection may
cause the sample to underrepresent clusters with unusually low X-ray
luminosities for their masses.  This incompleteness almost certainly
causes some of the deviations from the theoretical mass functions seen
at masses below $\times 10^{14} \msun$.  However, provided that the
scatter in the $L_X-M_{200}$ relation is not too large, this
incompleteness should be corrected for in the 1/$V_{max}(L_X)$
weighting.  Also, our redshift cutoff of $z<0.1$ likely increases the
completeness because many of the lower-mass clusters are in the Abell
catalog and some of the X-ray cluster catalogs explicitly searched for
X-ray emission around Abell clusters.  This effect likely dominates
the competing effect that low-redshift, lower-mass clusters may have
lower X-ray surface brightnesses.

Even if the above effects all combine to bias our mass function to
erroneously low number densities, we estimate a maximum incompleteness of a
factor of two.  
Note that such a large incompleteness would yield best-fit parameters
$\Omega_m$=0.33 and $\sigma_8$=0.87, outside the 99.7\% confidence
contour of our standard estimate.  The best-fit contours computed
assuming the NORAS completeness of 52\% for the entire CIRS sample
(i.e., approximately doubling the number densities) produces contours
shifted towards larger values of $\Omega_m$ and, to a lesser degree,
towards larger values of $\sigma_8$.  Realistically, the
incompleteness is probably larger at smaller masses.  Correcting for
such a mass-dependent incompleteness would tend to decrease the
inferred estimate of $\sigma_8$ with a much smaller corresponding
increase in $\Omega_m$.

\subsubsection{Mass Estimation and Velocity Segregation}

\begin{figure}
\figurenum{11}
\epsscale{0.8}
\plotone{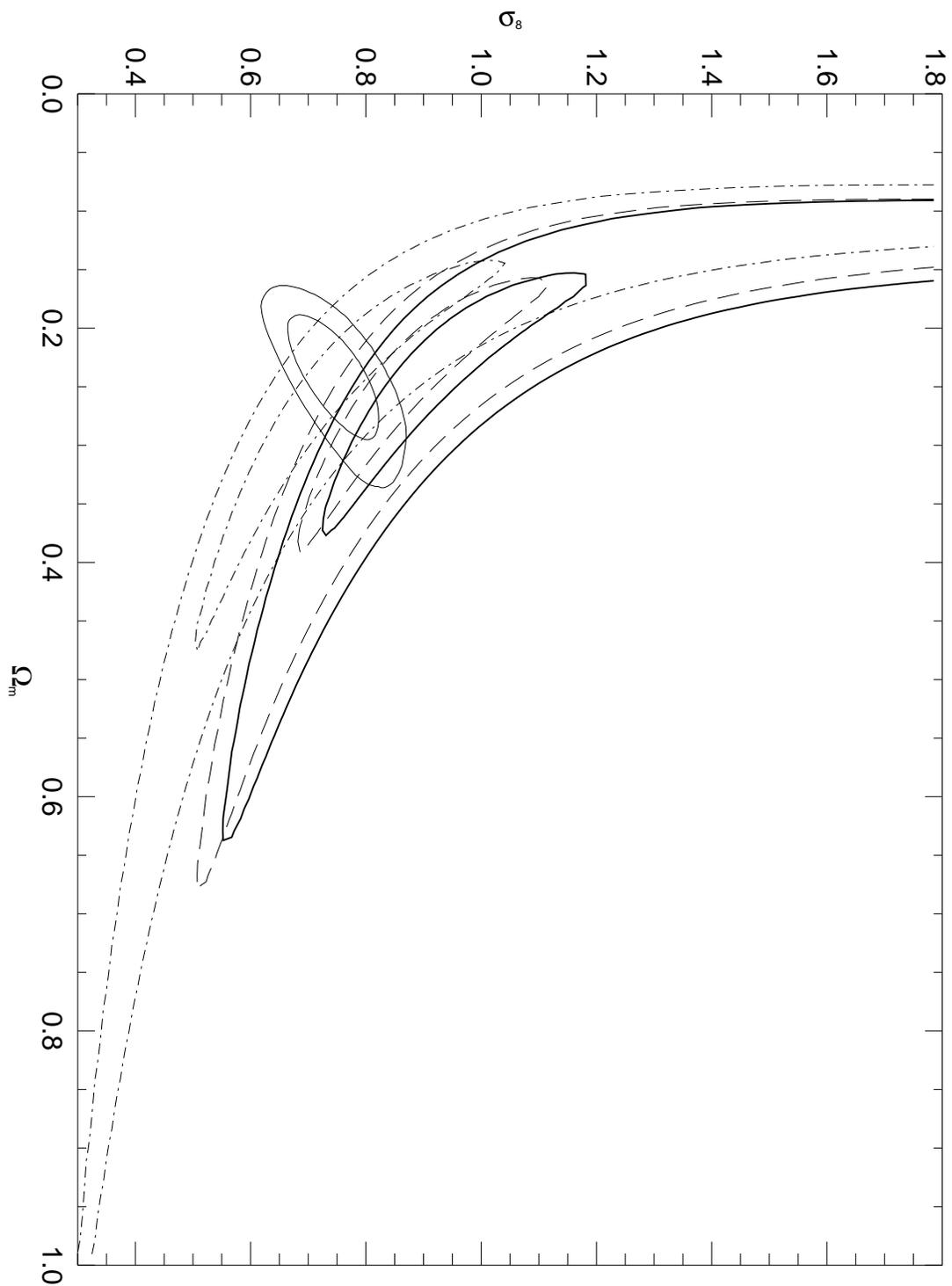}
\caption{\label{omsigsyssigm} 
Cosmological constraints from the CIRS virial mass function.  Solid
contours show 1-3$\sigma$ confidence levels for $\Omega_m$ and
$\sigma_8$.  Dashed and dash-dotted contours show the 1-3$\sigma$
confidence levels from adopting a mean uncertainty 2x and 5x the
statistical uncertainty respectively.  The solid contours 
extending to the lower 
left show the 68\% and 95\% confidence levels from WMAP3.}
\end{figure}

The largest potential systematic effect we find is artificially
increasing the uncertainty in the virial mass estimates.  As the
assumed uncertainty increases, the convolved mass function is more
dramatically flattened relative to the true mass function.  Figure
\ref{omsigsyssigm} shows the effects of increasing the virial mass
uncertainty $\sigma_{\mbox{log}M}$ by a factor of two (26\%
uncertainty, dashed contours) and by a factor of five (factor of two
uncertainty, dash-dotted contours) without using the $V_{max}$
weighting of Equation \ref{vmaxweight}.  We find that artificially
increasing the mass uncertainty by a factor of five and eliminating
the $V_{max}$ weighting is sufficient to bring the CIRS mass function
into agreement with the WMAP3 parameters, while the change is much
smaller when the $V_{max}$ weighting is included.  The effect of mass
uncertainties is much more severe for the poorly sampled systems of
\citet{eke05} and require detailed corrections from mock catalogs.
Because the CIRS clusters are much more densely sampled, the effect
should be much smaller for the CIRS mass function.  Indeed, Figure
\ref{omsignew4} shows that the constraints from the CIRS mass function
agree well with those of \citet{eke05}.  

We investigate two potential sources of systematic error caused by the
mass range of our sample.  Because the concentration $c$ is a weak
function of mass ($c\propto M^{-0.13}$), this may affect the
comparison of the mass function via the conversion from $M_{180b}$ to
$M_{200}$.  We refit the mass function using the relation of
\citet{bullock01} and find essentially the same results as for fixed
$c$.  Note that the CIRS sample showed little evidence of even a weak
trend of c with mass \citep[][but see Vikhlinin et al.~2006]{cirsi}.
Second, the uncertainty in the estimated mass may depend on the mass
of the system.  Less massive systems are expected to contain fewer
galaxies and thus may have larger statistical uncertainties.  We find
such a trend in the CIRS virial masses and redo the fits to the mass
function assuming an uncertainty that declines with mass according to
$\sigma_{\mbox{log}M}=0.34-0.02 \mbox{log}M$ in the convolution of the
theoretical mass functions.  The resulting confidence levels have
nearly identical shapes and locations to our default contours.

The issue of velocity segregation between galaxies and dark matter
remains an unresolved problem
\citep{kauffmann1999a,kauffmann1999b,diemand04,faltenbacher05}.
Velocity bias might be significant in the centers of clusters, where
the galaxies have experienced dynamical friction.  Dynamical friction
should produce a smaller velocity dispersion for cluster galaxies than
dark matter
\citep{1999ApJ...516..530K,2003ApJ...590..654Y,2005MNRAS.361.1203V},
although they may undergo more frequent mergers (due to their smaller
velocities) or tidal disruption, resulting in a larger velocity
dispersion of the surviving cluster galaxies relative to the dark
matter \citep{1999ApJ...523...32C,diemand04,faltenbacher05}.  Note
also that the velocity bias may depend on the mass of the cluster
\citep{berlind03} or the luminosities of the tracer galaxies
\citep{2006MNRAS.366.1455S}.  An important caveat is that the subhalo
distribution of cluster galaxies in simulations does not match the
observed distributions of cluster galaxies: real cluster galaxies
approximately trace the dark matter profile
\citep[e.g.,][]{cye97,rines01a,bg03,katgert03,lokas03,lin04,cairnsii}, while 
simulated cluster galaxies are significantly antibiased in cluster
centers. This difference is likely caused by the resilience of the
stellar component of a subhalo (i.e., real galaxies) against tidal
disruption relative to the total subhalo in simulations, implying that
a better understanding of the difference between galaxies and
subhaloes is required for accurate models of velocity segregation
\citep[e.g.,][]{diemand04,faltenbacher06}.  For instance, the details
of velocity bias depend on the selection used to identify subhaloes,
e.g., whether subhalo mass is measured prior to or after a subhalo
enters the cluster \citep{faltenbacher06}.

The amount of velocity bias in simulations is
typically $\sim$10\% for cluster-size haloes, although disagreement
remains on whether this bias is positive or negative.  Velocity bias
of this size would lead to virial masses overestimated or
underestimated by $\sim$20\%.  Future work is clearly needed to
determine the nature and significance of velocity bias for cluster
mass estimates.  There is no clear evidence from observations in favor
of or against velocity bias in clusters, although the generally good
agreement between X-ray, lensing, and virial mass estimates suggests
that any velocity bias is not large \citep[$\lesssim$20\%;
e.g.,][]{girardi98,popesso04,diaferio05}.  Because significant
velocity bias would produce incorrect virial mass estimates, the
agreement between virial and caustic mass estimates is no guarantee
that the caustic mass estimates are accurate.  However, the mechanisms
which cause velocity bias are less effective in the outskirts of
clusters, so the caustic technique should be less affected by velocity
bias than estimates based on Jeans' analysis.  

\citet{biviano06} test the robustness of virial mass estimates 
in a cosmological hydrodynamic simulation.  They find that virial mass
estimates are generally reliable for densely sampled clusters,
yielding overestimates of $\approx$10\% for systems with more than 60
members.  In general, they find that the most significant problem with
estimating virial masses is adequate removal of interlopers.  The use
of the caustic technique, a relatively conservative membership
definition, should mitigate this problem for the CIRS sample.


If one is willing to take the WMAP results as a prior, then one can
estimate the sense and degree of velocity segregation by requiring
that the CIRS mass function reproduce the predictions of WMAP.  The
simplest implementation of this method is to assume that the virial
masses are biased by a (mass-independent) factor S such that the true
cluster mass is given by $M_{200,true}=SM_{200,obs}$.  If we use the
WMAP1 parameters as a prior, we obtain $S=1.14\pm0.12$, implying a
velocity segregation of
$\sigma_{gxy}/\sigma_{DM}\approx$0.94$\pm$0.05.  Taking the WMAP3
parameters as a prior, we obtain $S=0.61\pm0.05$, implying a velocity
segregation of $\sigma_{gxy}/\sigma_{DM}\approx$1.28$\pm$0.06.
\citet{seljak06} combined constraints from WMAP3 and a variety of
other techniques (Lyman-$\alpha$ forest, SDSS galaxy clustering, SDSS
baryonic acoustic oscillations, supernovae, and smaller-scale CMB
measurements) to improve constraints on cosmological parameters.  They
find that the best-fit estimate of $\sigma_8$ increases to
$\sigma_8=0.847\pm0.022$ with essentially no change in
$\Omega_m=0.235\pm0.008$.  If we repeat the above analysis using these
parameters as priors, we find $S=0.83\pm0.08$ and
$\sigma_{gxy}/\sigma_{DM}\approx$1.10$\pm$0.05.  Thus, an independent
measurement of cosmological parameters can constrain the detailed
properties of cluster galaxies \citep{eke05,berlind06}.

Blue and red galaxies in clusters have dramatically different
spatial distributions.  By using all galaxies within the caustics to
estimate virial masses, we may overestimate $M_{200}$.  We recalculate
the virial mass profiles using only red galaxies, where we define red
galaxies as those with ($g$-$r$)$\geq$-0.05($M_r$+20)+0.7 where $M_r$
is the absolute magnitude in $r$ band and $g$ and $r$ are Petrosian
magnitudes.  We find that there is very little difference in the
resulting virial mass estimates and, as a result, little change in the
estimated mass function or cosmological parameters (Figure
\ref{omsigsysmass}).  This result differs from the predictions of
\citet{biviano06} that the virial masses estimated with all galaxies
are biased high by a factor of $\sim$1.40, while the virial masses
computed with red galaxies only are approximately unbiased
($M_{vir,red}\approx1.01M_{vir,true}$).  We suggest that this
difference is due to the difference in the contrast of caustics in
observations and in simulations \citep{cairnsi,cirsi}.  A more
detailed study of the velocity distribution of red and blue galaxies
in and around clusters suggests that intrinsic differences in the
velocity dispersions of the two populations are substantially reduced
when using the caustic technique to select members (Rines \& Diaferio
in prep.).  This study also shows that, when using the caustic
technique, the inclusion or exclusion of blue galaxies produces little
change in the velocity dispersion of the total population (in part
because blue galaxies are rare in clusters).

Most of the CIRS clusters are sampled only once by SDSS.  Thus, the
fiber placement constraints of SDSS (no two fibers can be placed
closer than 60\arcs) causes non-uniform sampling of the cluster
galaxies.  \citet{biviano06} show that non-uniform sampling of cluster
galaxies by multislit or multifiber surveys may bias virial mass
estimates (the cluster core is undersampled so the virial radius is
biased).  Unfortunately, few clusters have sufficient coverage in both
SDSS and elsewhere to test this effect directly.  Followup
spectroscopy of the CIRS clusters to complete the spectroscopic
samples from SDSS would test the significance of this effect.

\subsubsection{Peculiar Velocities}

We make no corrections for peculiar velocities of the CIRS clusters
because no reliable peculiar velocity estimates exist for most of the
sample.  The closest clusters will be most strongly affected by this
approach, because their distance estimates are most sensitive to a
given correction for peculiar velocity.  Our minimum redshift of
$z$$>$0.02 therefore significantly reduces this uncertainty.

We estimate the magnitude of this effect by adding a random peculiar
velocity from a Gaussian distribution of width 400$\kms$
\citep{sheth01} to each cluster redshift and recomputing the cluster
mass function and cosmological constraints.  Several trials using this
procedure find confidence contours for ($\Omega_m$,$\sigma_8$) that
agree well with our default contours.  We conclude that peculiar
velocities have a small impact on our results, although extensive
Monte Carlo simulations could better quantify the resulting systematic
uncertainty.

\subsubsection{Large-Scale Structure}

\begin{figure}
\figurenum{12}
\plotone{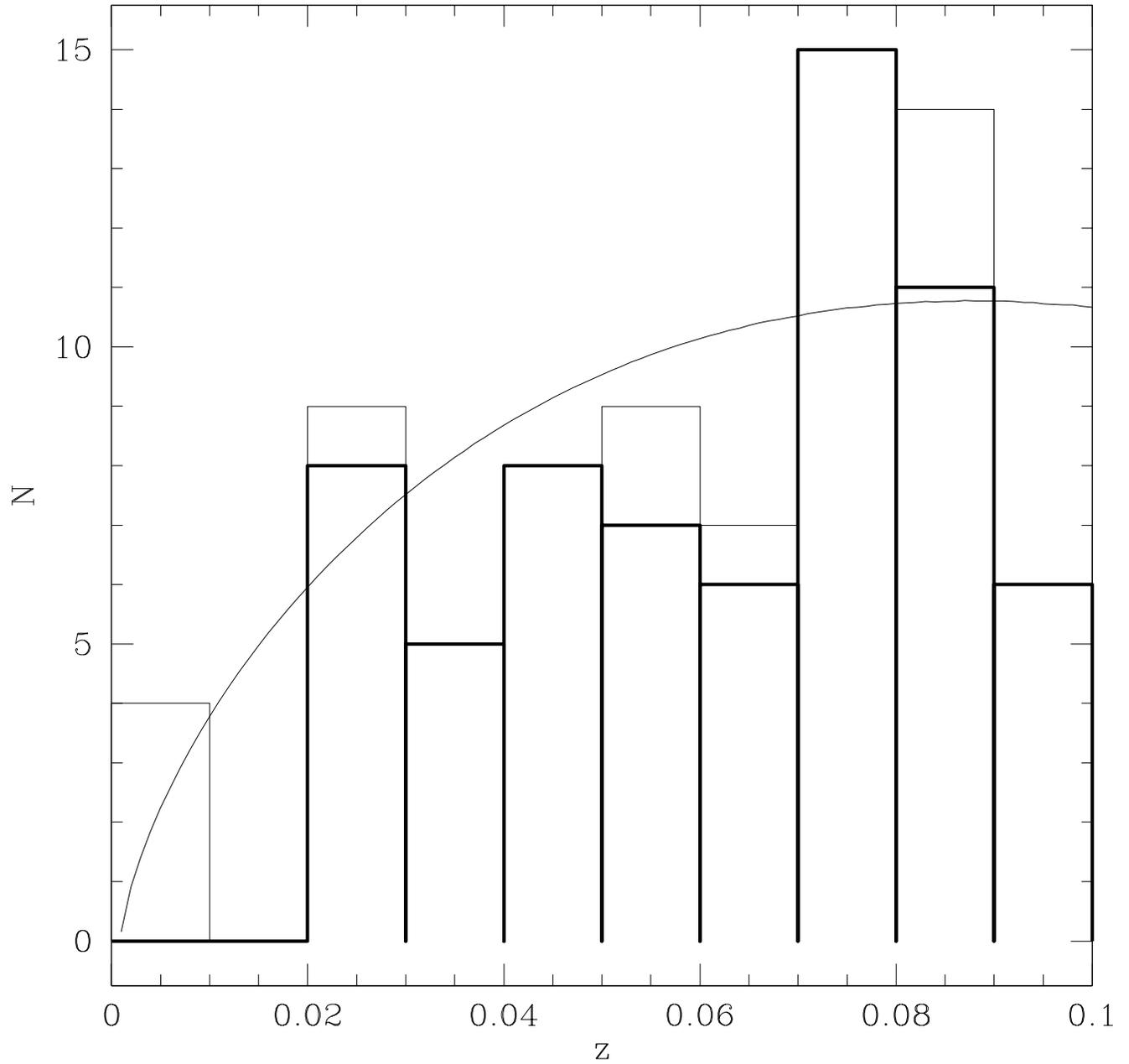}
\caption{\label{mfnz} 
Redshift histogram of the CIRS mass function sample (thick black
line). The thin black histogram shows the redshift histogram of the
``maximal'' CIRS sample including cluster pairs and systems at low
redshift.  The solid curve shows the expected number of clusters for
the X-ray luminosity function of \citet{bohringer02}.  }
\end{figure}

Large-scale structure could have a significant impact on our mass
function if the scale of overdensities is large compared to the volume
of the survey.  The CIRS sky coverage is that of the spectroscopic
survey of SDSS DR4.  This means that CIRS does not include the Coma
cluster or much of the Great Wall \citep{gh89}, but it does include
the Sloan Great Wall \citep{gott05}.  Because of the redshift limit,
the maximum volume probed is $\sim 10^8 h^{-3}Mpc^3$ (Figure
\ref{vmax}).  Figure \ref{mfnz} shows a histogram of the redshift
ditribution of the CIRS mass function sample.  The solid curve in
Figure \ref{mfnz} shows the expected redshift distribution for the
CIRS sample given the X-ray luminosity function of
\citet{bohringer02}.  There is no strong evidence for large-scale
variations in this histogram. It is difficult to determine whether the
CIRS sample contains an overdense or underdense sample of the
universe.  The eventual release of SDSS DR5 and the SDSS-II Legacy
Survey will approximately double the spectroscopic area covered, so
those data will provide a good test of the representativeness of the
CIRS sample.

\subsubsection{Scaling Relations}

An alternative method of estimating mass functions is to measure
simple observables such as X-ray luminosity or galaxy richness and
then apply scaling relations to estimate cluster masses.  Although
this method has been applied extensively
\citep{pierpaoli01,seljak02,pierpaoli03}, the reliability of the mass
function and/or cosmological constraints is limited by the reliability
of the scaling relations.  Scaling relations often have large scatter
\citep{popesso04} and they are difficult to calibrate
\citep{emn96,stanek06}.  We recalculate the CIRS mass function using
the observed X-ray luminosities and two scaling relations for
$L_X-M_{200}$, one based on virial masses \citep{cirsi} and one based
on masses from X-ray temperatures \citep{popesso04}.  The resulting
mass function for the latter relation contains many fewer systems at
M$\lesssim$10$^{14}\msun$ than the CIRS mass function (Figure
\ref{mfn}).  This difference seems to be caused by a difference in the
scaling relations determined by \citet{popesso04} and
\citet{hiflugcs}.  \citet{cirsi} noted that their $M_{vir}-L_X$
relation was offset from that of
\citet{popesso04}, although the $M-T_X$ relations agree.  The
$L_X-M_{200}$ of \citet{hiflugcs} and \citet{cirsi} agree with each
other, and so do the resulting estimates of the mass function.  We
therefore caution that great care and judgment must be taken when
using scaling relations to measure the cluster mass function.

\subsubsection{Sample Selection}

There are several methods of identifying clusters.  Four techniques
are used to identify clusters: galaxy overdensities (either in
projection or in redshift space), identifying extended X-ray sources,
identifying peaks in gravitational lensing maps, and identifying
Sunyaev-Zeldovich decrements in maps of the cosmic microwave
background.  Lensing and SZ surveys are both very promising but have
yet to produce large, well-selected samples.

Identification of galaxy overdensities has proved very successful, but
this method is subject to projection effects, especially when
performed without redshift information. \citet{eke05} study the mass
function from an optically-selected group catalog containing systems
with as few as two members.  They justify this procedure by extensive
comparison to mock catalogs, but the resulting cosmological
constraints then have additional systematic uncertainty due to any
differences between the mock catalogs and the observations.

\citet{popesso06} describe a targeted search for X-ray emission 
around Abell clusters in SDSS.  They find that many Abell clusters
with apparently Gaussian velocity distributions lie far below the
$L_X-M$ relation of X-ray selected clusters, although they lie on the
same $L_{opt}-M$ relation.  If these X-ray faint clusters are indeed
very massive, then the constraints on $\Omega_m$ and $\sigma_8$ from
the CIRS mass function are underestimates.  Note, however, that these
changes would lead to larger disagreements with other estimates
(Figure \ref{omsig}), unless either velocity segregation is large
(and therefore the masses are overestimated) or the true uncertainty
in the measured virial masses is much larger than the statistical
uncertainty (or some combination of these effects).

It would be very instructive to compare the CIRS mass function to a
cluster mass function constructed with optical cluster catalogs from
SDSS \citep[e.g.,][]{miller05}, but this analysis lies outside the
scope of this work.  Detailed comparisons of cluster catalogs selected
with different techniques \citep[e.g.,][]{donahue02} are important for
using clusters to constrain cosmological parameters.

\subsubsection{Uncertainties in the Predicted Mass Function}

Most early attempts to measure cosmological parameters from the
observed mass function of clusters used Press-Schechter formalism
\citep{ps74}.  In recent years, however, numerical simulations have
achieved sufficient resolution to directly measure the mass function
of dark matter haloes in cosmologically representative volumes.
\citet{sheth99} proposed a modified form of the Press-Schechter
relation that provides a better match to simulations.
\citet{jenkins01} provide a third form of this relation which provides
a good match to simulations of the Virgo consortium.  We use their
fitting formula for the mass function along with the power spectrum of
\citet{bardeen86}.  \citet{hiflugcs} notes that the best-fit
cosmological parameters are unchanged if one uses the power spectrum
of \citet{eisenstein99} instead.
\citet{warren06} use 16 nested simulations to test the effects of numerical
resolution on the mass function.  Although they find significant
differences from the \citet{jenkins01} relation for extreme masses
(M$<$10$^{13}\msun$ and M$>2\times$10$^{15}\msun$), the
\citet{jenkins01} relation provides a good approximation over the mass
range considered here.  The constraints in the ($\Omega_m$,$\sigma_8$)
plane obtained by using the Press-Schechter or Sheth-Tormen form are
nearly identical to those obtained with the \citet{jenkins01}
relation.

\citet{white02} discusses the importance of carefully defining the
mass of a halo when comparing theoretical and observed mass functions.
In particular, the universality of the mass function depends on the
algorithm used to measure the mass.  \citet{white02} concludes that
the relation of Equation (9) of \citet{jenkins01} provides a good match
to the mass measured within a spherical overdensity of 180 times the
{\it background} (not critical) density.  Both \citet{white02} and
\citet{hu03} recommend computing the mass function for a given set of
cosmological parameters using the 180b definition of mass and then
using an NFW profile with concentration $c_{200}$=5 to convert the
observed and theoretical mass scales to a common scale.  Here we
convert the mass scale of the theoretical mass function so that the
models can be shown against the observations more easily.  There
appears to be consensus in the literature that the systematic
uncertainties in the predicted mass function for different
cosmological models are of the order 15\%
\citep{white02,evrard02b,hu03,stanek06}.  \citet{white02} suggests that
higher accuracy may require a suite of cosmological simulations run
with different values of $\Omega_m$ and $\sigma_8$ with subsequent
analysis using a variety of algorithms for defining the mass of a
halo.

Some authors have noted that the most massive clusters measure
fluctuations on a scale larger than 8$\Mpc$.  In particular,
\citet{pierpaoli01} note that clusters with temperatures of 6.5 keV
sample fluctuations on a scale of $\approx$14$\Mpc$.  The CIRS sample
probes lower mass systems than many previous studies: the median mass
of the sample is 2$\times$10$^{14}\msun$, corresponding to a scale of
8.3$\Mpc$ for $\Omega_m$=0.3.  Thus, we constrain $\sigma_8$ directly.
Virial masses have an advantage over X-ray masses in this respect
because the properties of the ICM for clusters with
M$\sim$10$^{14}\msun$ may be strongly affected by feedback from
supernovae and AGN \citep[e.g.,][]{loewenstein00}, and the ICM in
groups may not be in hydrostatic equilibrium.

\section{Velocity Dispersion Function}

\begin{figure}
\figurenum{13}
\plotone{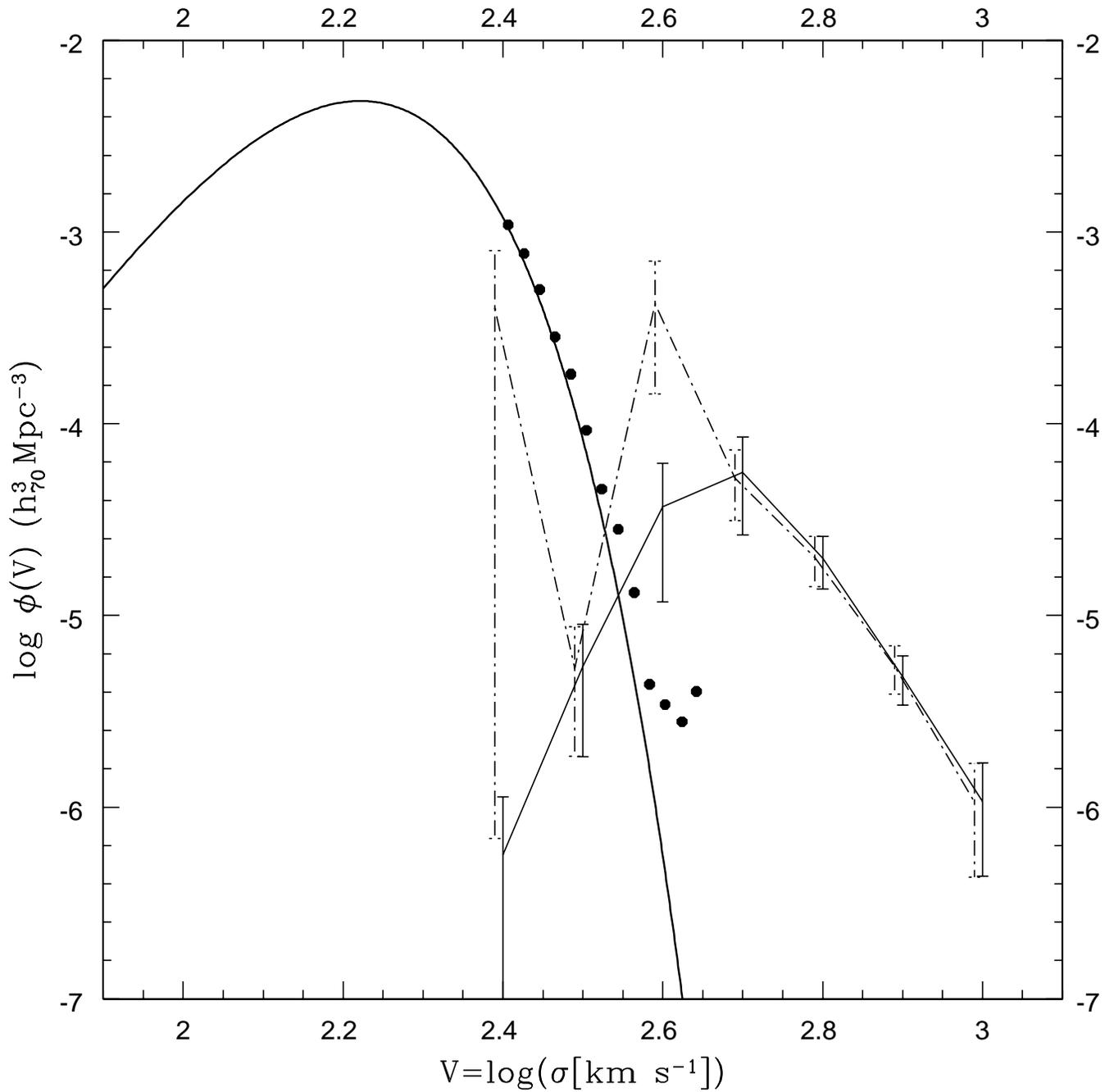}
\caption{\label{dsigmafn} 
The velocity dispersion function of the CIRS sample.  The solid curve
shows the best-fit velocity dispersion function of early-type galaxies
in SDSS and the black dots indicate the deviations from this fit at
log$\sigma$$\gtrsim$2.4 \citep{mitchell05}.  The dash-dotted line is the
velocity dispersion function of the ``maximal'' CIRS sample including
cluster pairs and low-redshift systems. }
\end{figure}

The abundance of systems as a function of velocity dispersion is an
important input to models of gravitational lensing.  \citet{sheth03}
and \citet{mitchell05} compute the velocity dispersion function of
early-type galaxies in SDSS.  One might expect a smooth transition
between the velocity dispersion function of galaxies and the velocity
dispersion function of groups and clusters \citep{shimasaku93}.
\citet{1993AJ....106.1301Z} found the cluster velocity dispersion
function for systems identified in the CfA Redshift Survey \citep[see
also][]{pisani03}.  They found good agreement with the velocity
dispersion function implied by the X-ray temperature function.  Figure
\ref{dsigmafn} compares the velocity dispersion functions of these two
types of systems.  The cluster velocity dispersion function shows a
deficit of low-$\sigma$ systems relative to the number needed to
produce a smooth function.  The dashed line in Figure
\ref{dsigmafn} shows the velocity dispersion function for our
``maximal'' mass function sample including cluster pairs and
low-redshift systems.  This estimate is more consistent with a
smooth transition between dark matter haloes containing one versus
many galaxies, although the two types of systems still do not join
smoothly.  A careful study of the abundance of galaxy groups that span
the intermediate mass and velocity dispersion range would shed light
on this transition.  Earlier studies
\citep{1993AJ....106.1301Z,mazure96,girardi00b,heinamaki03,pisani03} were
generally limited to small, nearby volumes, so cosmic variance will
continue to be an important limitation to these efforts.  The much
larger samples available with 2dFGRS and SDSS should alleviate this
problem \citep{martinez02b,eke05}.

\begin{figure}
\figurenum{14}
\plotone{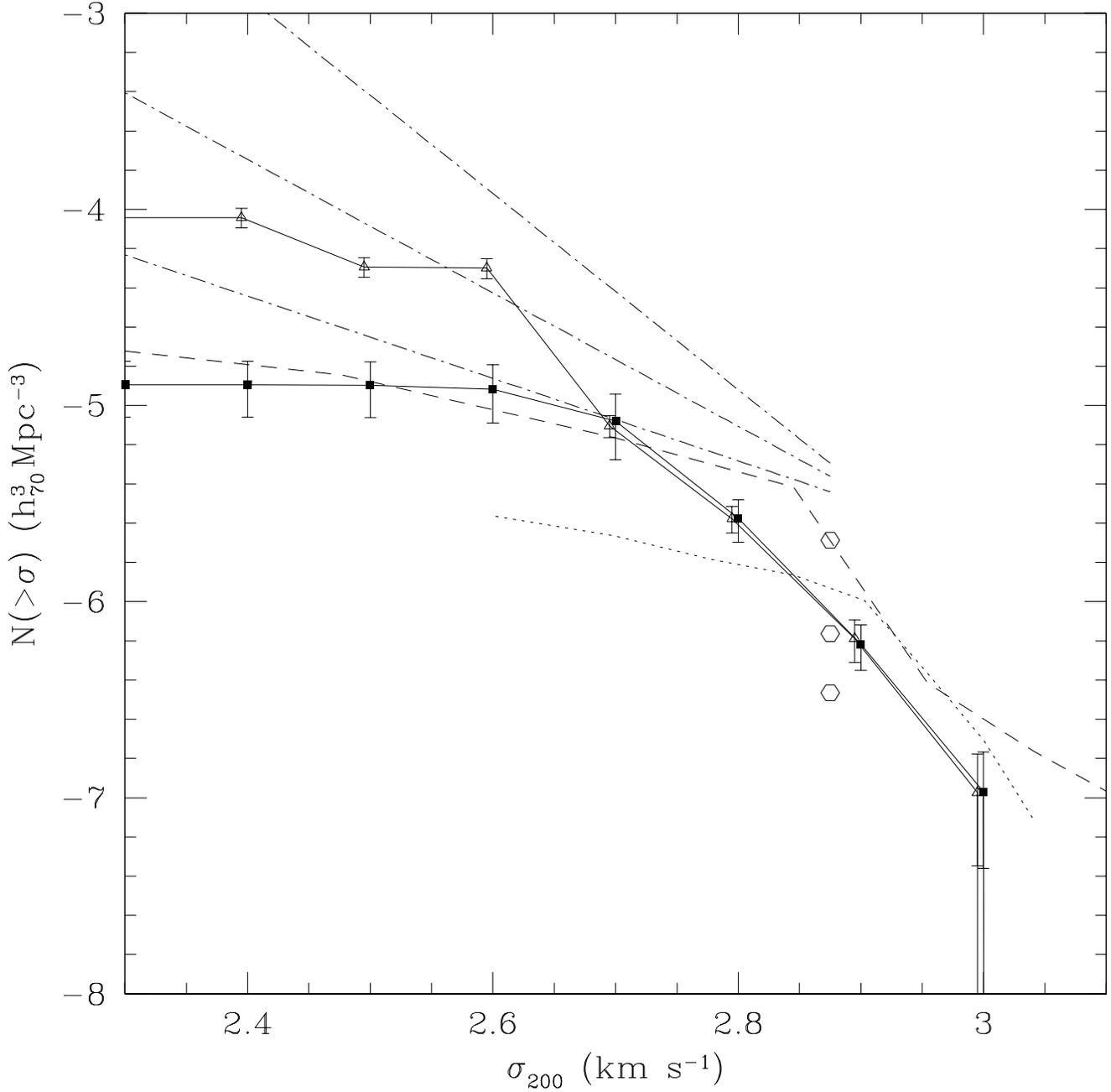}
\caption{\label{sigmafn} 
The cumulative velocity dispersion function of the CIRS sample
(squares).  The triangles show the velocity dispersion function of the
``maximal'' CIRS sample including cluster pairs and low-redshift
systems.  The dash-dotted lines show the estimate and 95\% confidence
range from \citet{pisani03}, and circles show their compilation of
previous results at $\sigma$=750$\kms$.  Dashed and dotted lines show
estimates from \citet{1993AJ....106.1301Z} and \citet{mazure96}.}
\end{figure}

Figure \ref{sigmafn} shows the cumulative velocity dispersion function
of the CIRS sample.  Our estimate is lower than that of
\citet{pisani03} from the CfA2 Redshift Survey, but they note that
their estimate is larger than most previous estimates, a result they
attribute to cosmic variance due to the presence of the Great Wall in
the survey.  The CIRS velocity dispersion function agrees with these
earlier estimates \citep[e.g.,][]{1993AJ....106.1301Z,mazure96}.
However, \citet{casagrande06} analyze the properties of groups in the
Updated Zwicky Catalog (UZC) and find similar results to
\citet{pisani03}.

\section{Discussion and Future Work}

We use the Fourth Data Release of the Sloan Digital Sky Survey to
measure the mass function of galaxy clusters in the local universe.
We select clusters based on their X-ray flux and a redshift limit
($z<0.1$).  X-ray selection has the crucial advantage over optical
selection of yielding a well-defined selection function.  We measure
cluster masses from the dynamics of their member galaxies using both
the virial theorem and the caustic technique.  The resulting mass
functions agree well with previous determinations using X-ray
properties and the mass-richness relation.  We conclude that the
cluster mass function is measured robustly in the local universe.  We
compare this mass function to predictions of various cosmological
models to obtain constraints in the ($\Omega_m$,$\sigma_8$) plane.
With the CIRS virial mass function (and assuming $h$=0.7), we estimate
$\Omega_m$=$0.24^{+0.14}_{-0.09}$ and
$\sigma_8$=$0.92^{+0.24}_{-0.19}$, or $\sigma_8$=$0.84\pm$0.03 when
holding $\Omega_m$=0.3 fixed.

The CIRS virial mass function agrees well with the WMAP1 and lies
slightly above the WMAP3 results, although still within the 95\%
confidence levels.  This result contrasts slightly with
\citet{reiprich06}, who finds that the HIFLUGCS mass function lies
much closer to WMAP3 than WMAP1.  However, other investigators have
found mass functions using X-ray data that lie closer to the CIRS mass
function, and there remain open questions about the systematic
uncertainties in the X-ray mass estimates \citep{rasia05,stanek06}.
Virial masses have the following advantages: they measure the masses
at larger radii, where less extrapolation is required to compare with
theoretical results, they are less sensitive to the complex physics in
the centers of clusters, and they can be applied to lower mass systems
for which nongravitational physics greatly complicates X-ray mass
estimates, thus eliminating any uncertainty due to possible scale
dependence of the estimate of $\sigma_8$ \citep[the clusters of, e.g.,][sample
fluctuations on larger scales, $\sigma_{14}$, which must then be
converted to $\sigma_8$]{pierpaoli01}.  We investigate several
potential sources of systematic error and conclude that the most
significant effect would be a large underestimate of the uncertainties
in the virial mass estimates.  Recent investigations with numerical
simulations \citet{biviano06} suggest that the scatter in virial mass
estimates of well-sampled clusters is smaller than the amount required
to bring our results into agreement with the WMAP3 parameters,
although further confirmation of this result would be useful.

Although obtaining virial mass estimates from galaxy dynamics for
large numbers of clusters requires much observational effort, the
reward is an independent measurement of the mass function and a
powerful confirmation of the current cosmological model.
This work further demonstrates the power of the mass function as a
fundamental constraint on cosmological parameters.  The SDSS-II Legacy
Survey should substantially enlarge the sample of local clusters and
reduce both statistical and systematic uncertainties in the mass
function.  

The CIRS mass function provides a critical normalization of the mass
function in the nearby universe.  An accurate normalization of the
local mass function is a prerequisite for future studies of the
evolution of the mass function.  Future determinations of the mass
function at higher redshifts will directly probe the growth rate of
structure and provide further constraints on cosmological parameters
\citep[e.g.,][]{haiman01,hu03b,vikhlinin03,henry04,majumdar04,vv04}.  
We conclude that a large spectroscopic program to measure virial
masses of an X-ray selected sample at moderate redshift would provide
an independent measurement of the evolution of the mass function.

\acknowledgements

We thank Adrian Jenkins for providing his software for calculating the
mass function.  We thank Margaret Geller for helpful discussions and
suggestions.  We thank Michael Hudson and the anonymous referee for
helpful suggestions which improved the presentation of this paper.  We
thank the Smithsonian Astrophysical Observatory for hosting KR as a
visitor for part of this project and for use of computational
facilities.  Funding for the Sloan Digital Sky Survey (SDSS) has been
provided by the Alfred P. Sloan Foundation, the Participating
Institutions, the National Aeronautics and Space Administration, the
National Science Foundation, the U.S. Department of Energy, the
Japanese Monbukagakusho, and the Max Planck Society. The SDSS Web site
is http://www.sdss.org/.  The SDSS is managed by the Astrophysical
Research Consortium (ARC) for the Participating Institutions. The
Participating Institutions are The University of Chicago, Fermilab,
the Institute for Advanced Study, the Japan Participation Group, The
Johns Hopkins University, the Korean Scientist Group, Los Alamos
National Laboratory, the Max-Planck-Institute for Astronomy (MPIA),
the Max-Planck-Institute for Astrophysics (MPA), New Mexico State
University, University of Pittsburgh, University of Portsmouth,
Princeton University, the United States Naval Observatory, and the
University of Washington.

\bibliographystyle{apj}
\bibliography{rines}

\begin{thebibliography}{137}
\expandafter\ifx\csname natexlab\endcsname\relax\def\natexlab#1{#1}\fi

\bibitem[{{Abell} {et~al.}(1989){Abell}, {Corwin}, \& {Olowin}}]{aco1989}
{Abell}, G.~O., {Corwin}, H.~G., \& {Olowin}, R.~P. 1989, \apjs, 70, 1

\bibitem[{{Adelman-McCarthy} {et~al.}(2006)}]{dr4}
{Adelman-McCarthy}, J. {et~al.} 2006, \apjs, 162, 38

\bibitem[{{Allen} {et~al.}(2003){Allen}, {Schmidt}, {Fabian}, \&
  {Ebeling}}]{allen03}
{Allen}, S.~W., {Schmidt}, R.~W., {Fabian}, A.~C., \& {Ebeling}, H. 2003,
  \mnras, 342, 287

\bibitem[{{B{\" o}hringer} {et~al.}(2000)}]{noras}
{B{\" o}hringer}, H. {et~al.} 2000, \apjs, 129, 435

\bibitem[{{B{\" o}hringer} {et~al.}(2001)}]{2001A&A...369..826B}
---. 2001, \aap, 369, 826

\bibitem[{{B{\" o}hringer} {et~al.}(2004)}]{2004A&A...425..367B}
---. 2004, \aap, 425, 367

\bibitem[{{Bahcall} \& {Bode}(2003)}]{bahcall03b}
{Bahcall}, N.~A. \& {Bode}, P. 2003, \apjl, 588, L1

\bibitem[{{Bahcall} \& {Cen}(1993)}]{bahcall93}
{Bahcall}, N.~A. \& {Cen}, R. 1993, \apjl, 407, L49

\bibitem[{{Bahcall} {et~al.}(2000){Bahcall}, {Cen}, {Dav{\' e}}, {Ostriker}, \&
  {Yu}}]{bahcall2000}
{Bahcall}, N.~A., {Cen}, R., {Dav{\' e}}, R., {Ostriker}, J.~P., \& {Yu}, Q.
  2000, \apj, 541, 1

\bibitem[{{Bahcall} {et~al.}(2003)}]{bahcall03a}
{Bahcall}, N.~A. {et~al.} 2003, \apj, 585, 182

\bibitem[{{Bardeen} {et~al.}(1986){Bardeen}, {Bond}, {Kaiser}, \&
  {Szalay}}]{bardeen86}
{Bardeen}, J.~M., {Bond}, J.~R., {Kaiser}, N., \& {Szalay}, A.~S. 1986, \apj,
  304, 15

\bibitem[{{Benatov} {et~al.}(2006){Benatov}, {Rines}, {Natarajan}, {Nagai}, \&
  {Kravtsov}}]{benatov06}
{Benatov}, L., {Rines}, K.~J., {Natarajan}, P., {Nagai}, D., \& {Kravtsov}, A.
  2006, \mnras, in press (astro-ph/0605105)

\bibitem[{{Berlind} {et~al.}(2003)}]{berlind03}
{Berlind}, A.~A. {et~al.} 2003, \apj, 593, 1

\bibitem[{{Berlind} {et~al.}(2006)}]{berlind06}
---. 2006, ArXiv Astrophysics e-prints

\bibitem[{{Biviano} \& {Girardi}(2003)}]{bg03}
{Biviano}, A. \& {Girardi}, M. 2003, \apj, 585, 205

\bibitem[{{Biviano} \& {Katgert}(2004)}]{biviano04}
{Biviano}, A. \& {Katgert}, P. 2004, \aap, 424, 779

\bibitem[{{Biviano} {et~al.}(2006){Biviano}, {Murante}, {Borgani}, {Diaferio},
  {Dolag}, \& {Girardi}}]{biviano06}
{Biviano}, A., {Murante}, G., {Borgani}, S., {Diaferio}, A., {Dolag}, K., \&
  {Girardi}, M. 2006, ArXiv Astrophysics e-prints

\bibitem[{{Blanton} {et~al.}(2003)}]{blanton03}
{Blanton}, M.~R. {et~al.} 2003, \apj, 592, 819

\bibitem[{{B{\"o}hringer} {et~al.}(2002)}]{bohringer02}
{B{\"o}hringer}, H. {et~al.} 2002, \apj, 566, 93

\bibitem[{{Borgani} {et~al.}(2001)}]{borgani01}
{Borgani}, S. {et~al.} 2001, \apj, 561, 13

\bibitem[{{Brada{\v c}} {et~al.}(2004){Brada{\v c}}, {Lombardi}, \&
  {Schneider}}]{bradac04}
{Brada{\v c}}, M., {Lombardi}, M., \& {Schneider}, P. 2004, \aap, 424, 13

\bibitem[{{Bullock} {et~al.}(2001)}]{bullock01}
{Bullock}, J.~S. {et~al.} 2001, \mnras, 321, 559

\bibitem[{{Carlberg} {et~al.}(1997){Carlberg}, {Yee}, \& {Ellingson}}]{cye97}
{Carlberg}, R.~G., {Yee}, H.~K.~C., \& {Ellingson}, E. 1997, \apj, 478, 462

\bibitem[{{Carlberg} {et~al.}(1996){Carlberg}, {Yee}, {Ellingson}, {Abraham},
  {Gravel}, {Morris}, \& {Pritchet}}]{cnoc96}
{Carlberg}, R.~G., {Yee}, H.~K.~C., {Ellingson}, E., {Abraham}, R., {Gravel},
  P., {Morris}, S., \& {Pritchet}, C.~J. 1996, \apj, 462, 32

\bibitem[{{Casagrande} \& {Diaferio}(2006)}]{casagrande06}
{Casagrande}, L. \& {Diaferio}, A. 2006, ArXiv Astrophysics e-prints

\bibitem[{{Col{\' i}n} {et~al.}(1999){Col{\' i}n}, {Klypin}, {Kravtsov}, \&
  {Khokhlov}}]{1999ApJ...523...32C}
{Col{\' i}n}, P., {Klypin}, A.~A., {Kravtsov}, A.~V., \& {Khokhlov}, A.~M.
  1999, \apj, 523, 32

\bibitem[{{Colless} {et~al.}(2001)}]{2df}
{Colless}, M. {et~al.} 2001, \mnras, 328, 1039

\bibitem[{{Cross} {et~al.}(2004){Cross}, {Driver}, {Liske}, {Lemon}, {Peacock},
  {Cole}, {Norberg}, \& {Sutherland}}]{2004MNRAS.349..576C}
{Cross}, N.~J.~G., {Driver}, S.~P., {Liske}, J., {Lemon}, D.~J., {Peacock},
  J.~A., {Cole}, S., {Norberg}, P., \& {Sutherland}, W.~J. 2004, \mnras, 349,
  576

\bibitem[{{Dahle}(2006)}]{dahle06}
{Dahle}, H. 2006, ArXiv Astrophysics e-prints

\bibitem[{{Diaferio}(1999)}]{diaferio1999}
{Diaferio}, A. 1999, \mnras, 309, 610

\bibitem[{{Diaferio} {et~al.}(2005){Diaferio}, {Geller}, \&
  {Rines}}]{diaferio05}
{Diaferio}, A., {Geller}, M.~J., \& {Rines}, K.~J. 2005, \apjl, 628, L97

\bibitem[{{Diemand} {et~al.}(2004){Diemand}, {Moore}, \& {Stadel}}]{diemand04}
{Diemand}, J., {Moore}, B., \& {Stadel}, J. 2004, \mnras, 352, 535

\bibitem[{{Donahue} {et~al.}(2002)}]{donahue02}
{Donahue}, M. {et~al.} 2002, \apj, 569, 689

\bibitem[{{Ebeling} {et~al.}(2000){Ebeling}, {Edge}, {Allen}, {Crawford},
  {Fabian}, \& {Huchra}}]{ebcs}
{Ebeling}, H., {Edge}, A.~C., {Allen}, S.~W., {Crawford}, C.~S., {Fabian},
  A.~C., \& {Huchra}, J.~P. 2000, \mnras, 318, 333

\bibitem[{{Ebeling} {et~al.}(1998){Ebeling}, {Edge}, {Bohringer}, {Allen},
  {Crawford}, {Fabian}, {Voges}, \& {Huchra}}]{bcs}
{Ebeling}, H., {Edge}, A.~C., {Bohringer}, H., {Allen}, S.~W., {Crawford},
  C.~S., {Fabian}, A.~C., {Voges}, W., \& {Huchra}, J.~P. 1998, \mnras, 301,
  881

\bibitem[{{Ebeling} {et~al.}(1996){Ebeling}, {Voges}, {Bohringer}, {Edge},
  {Huchra}, \& {Briel}}]{xbacs}
{Ebeling}, H., {Voges}, W., {Bohringer}, H., {Edge}, A.~C., {Huchra}, J.~P., \&
  {Briel}, U.~G. 1996, \mnras, 281, 799

\bibitem[{{Eisenstein} \& {Hu}(1999)}]{eisenstein99}
{Eisenstein}, D.~J. \& {Hu}, W. 1999, \apj, 511, 5

\bibitem[{{Eke} {et~al.}(2005){Eke}, {Baugh}, {Cole}, {Frenk}, \&
  {Navarro}}]{eke05}
{Eke}, V.~R., {Baugh}, C.~M., {Cole}, S., {Frenk}, C.~S., \& {Navarro}, J.~F.
  2005, ArXiv Astrophysics e-prints

\bibitem[{{Evrard} {et~al.}(1996){Evrard}, {Metzler}, \& {Navarro}}]{emn96}
{Evrard}, A.~E., {Metzler}, C.~A., \& {Navarro}, J.~F. 1996, \apj, 469, 494

\bibitem[{{Evrard} {et~al.}(2002)}]{evrard02b}
{Evrard}, A.~E. {et~al.} 2002, \apj, 573, 7

\bibitem[{{Fabian} {et~al.}(2006){Fabian}, {Sanders}, {Taylor}, {Allen},
  {Crawford}, {Johnstone}, \& {Iwasawa}}]{fabian06}
{Fabian}, A.~C., {Sanders}, J.~S., {Taylor}, G.~B., {Allen}, S.~W., {Crawford},
  C.~S., {Johnstone}, R.~M., \& {Iwasawa}, K. 2006, \mnras, 366, 417

\bibitem[{{Fabian} {et~al.}(2000)}]{fabian00}
{Fabian}, A.~C. {et~al.} 2000, \mnras, 318, L65

\bibitem[{{Fadda} {et~al.}(1996){Fadda}, {Girardi}, {Giuricin}, {Mardirossian},
  \& {Mezzetti}}]{1996ApJ...473..670F}
{Fadda}, D., {Girardi}, M., {Giuricin}, G., {Mardirossian}, F., \& {Mezzetti},
  M. 1996, \apj, 473, 670

\bibitem[{{Faltenbacher} {et~al.}(2005){Faltenbacher}, {Kravtsov}, {Nagai}, \&
  {Gottl{\"o}ber}}]{faltenbacher05}
{Faltenbacher}, A., {Kravtsov}, A.~V., {Nagai}, D., \& {Gottl{\"o}ber}, S.
  2005, \mnras, 358, 139

\bibitem[{{Faltenbacher} {et~al.}(2006){Faltenbacher}, {Kravtsov}, {Nagai}, \&
  {Gottl{\"o}ber}}]{faltenbacher06}
---. 2006, \mnras, submitted (astro-ph/0602197)

\bibitem[{{Geller} {et~al.}(1999){Geller}, {Diaferio}, \& {Kurtz}}]{gdk99}
{Geller}, M.~J., {Diaferio}, A., \& {Kurtz}, M.~J. 1999, \apjl, 517, L23

\bibitem[{{Geller} \& {Huchra}(1989)}]{gh89}
{Geller}, M.~J. \& {Huchra}, J.~P. 1989, Science, 246, 897

\bibitem[{{Girardi} {et~al.}(1998{\natexlab{a}}){Girardi}, {Borgani},
  {Giuricin}, {Mardirossian}, \& {Mezzetti}}]{girardi98b}
{Girardi}, M., {Borgani}, S., {Giuricin}, G., {Mardirossian}, F., \&
  {Mezzetti}, M. 1998{\natexlab{a}}, \apj, 506, 45

\bibitem[{{Girardi} \& {Giuricin}(2000)}]{girardi00b}
{Girardi}, M. \& {Giuricin}, G. 2000, \apj, 540, 45

\bibitem[{{Girardi} {et~al.}(1998{\natexlab{b}}){Girardi}, {Giuricin},
  {Mardirossian}, {Mezzetti}, \& {Boschin}}]{girardi98}
{Girardi}, M., {Giuricin}, G., {Mardirossian}, F., {Mezzetti}, M., \&
  {Boschin}, W. 1998{\natexlab{b}}, \apj, 505, 74

\bibitem[{{Gott} {et~al.}(2005){Gott}, {Juri{\'c}}, {Schlegel}, {Hoyle},
  {Vogeley}, {Tegmark}, {Bahcall}, \& {Brinkmann}}]{gott05}
{Gott}, J.~R.~I., {Juri{\'c}}, M., {Schlegel}, D., {Hoyle}, F., {Vogeley}, M.,
  {Tegmark}, M., {Bahcall}, N., \& {Brinkmann}, J. 2005, \apj, 624, 463

\bibitem[{{Haiman} {et~al.}(2001){Haiman}, {Mohr}, \& {Holder}}]{haiman01}
{Haiman}, Z., {Mohr}, J.~J., \& {Holder}, G.~P. 2001, \apj, 553, 545

\bibitem[{{Hamana} {et~al.}(2003)}]{hamana03}
{Hamana}, T. {et~al.} 2003, \apj, 597, 98

\bibitem[{{Hein{\"a}m{\"a}ki} {et~al.}(2003){Hein{\"a}m{\"a}ki}, {Einasto},
  {Einasto}, {Saar}, {Tucker}, \& {M{\"u}ller}}]{heinamaki03}
{Hein{\"a}m{\"a}ki}, P., {Einasto}, J., {Einasto}, M., {Saar}, E., {Tucker},
  D.~L., \& {M{\"u}ller}, V. 2003, \aap, 397, 63

\bibitem[{{Henry}(2004)}]{henry04}
{Henry}, J.~P. 2004, \apj, 609, 603

\bibitem[{{Henry} \& {Arnaud}(1991)}]{henry91}
{Henry}, J.~P. \& {Arnaud}, K.~A. 1991, \apj, 372, 410

\bibitem[{{Heymans} {et~al.}(2005)}]{heymans05}
{Heymans}, C. {et~al.} 2005, \mnras, 361, 160

\bibitem[{{Hoekstra} {et~al.}(2005){Hoekstra}, {Mellier}, {van Waerbeke},
  {Semboloni}, {Fu}, {Hudson}, {Parker}, {Tereno}, \& {Benabed}}]{hoekstra06}
{Hoekstra}, H., {Mellier}, Y., {van Waerbeke}, L., {Semboloni}, E., {Fu}, L.,
  {Hudson}, M.~J., {Parker}, L.~C., {Tereno}, I., \& {Benabed}, K. 2005, ArXiv
  Astrophysics e-prints

\bibitem[{{Hu}(2003)}]{hu03b}
{Hu}, W. 2003, \prd, 67, 081304

\bibitem[{{Hu} \& {Kravtsov}(2003)}]{hu03}
{Hu}, W. \& {Kravtsov}, A.~V. 2003, \apj, 584, 702

\bibitem[{{Jenkins} {et~al.}(2001){Jenkins}, {Frenk}, {White}, {Colberg},
  {Cole}, {Evrard}, {Couchman}, \& {Yoshida}}]{jenkins01}
{Jenkins}, A., {Frenk}, C.~S., {White}, S.~D.~M., {Colberg}, J.~M., {Cole}, S.,
  {Evrard}, A.~E., {Couchman}, H.~M.~P., \& {Yoshida}, N. 2001, \mnras, 321,
  372

\bibitem[{{Katgert} {et~al.}(2004){Katgert}, {Biviano}, \&
  {Mazure}}]{katgert03}
{Katgert}, P., {Biviano}, A., \& {Mazure}, A. 2004, \apj, 600, 657

\bibitem[{{Kauffmann} {et~al.}(1999{\natexlab{a}}){Kauffmann}, {Colberg},
  {Diaferio}, \& {White}}]{kauffmann1999a}
{Kauffmann}, G., {Colberg}, J.~M., {Diaferio}, A., \& {White}, S. D.~M.
  1999{\natexlab{a}}, \mnras, 303, 188

\bibitem[{{Kauffmann} {et~al.}(1999{\natexlab{b}}){Kauffmann}, {Colberg},
  {Diaferio}, \& {White}}]{kauffmann1999b}
---. 1999{\natexlab{b}}, \mnras, 307, 529

\bibitem[{{Klypin} {et~al.}(1999){Klypin}, {Gottl{\" o}ber}, {Kravtsov}, \&
  {Khokhlov}}]{1999ApJ...516..530K}
{Klypin}, A., {Gottl{\" o}ber}, S., {Kravtsov}, A.~V., \& {Khokhlov}, A.~M.
  1999, \apj, 516, 530

\bibitem[{{Lin} {et~al.}(2004){Lin}, {Mohr}, \& {Stanford}}]{lin04}
{Lin}, Y., {Mohr}, J.~J., \& {Stanford}, S.~A. 2004, \apj, 610, 745

\bibitem[{{Loewenstein}(2000)}]{loewenstein00}
{Loewenstein}, M. 2000, \apj, 532, 17

\bibitem[{{{\L}okas} \& {Mamon}(2003)}]{lokas03}
{{\L}okas}, E.~L. \& {Mamon}, G.~A. 2003, \mnras, 343, 401

\bibitem[{{{\L}okas} {et~al.}(2006){{\L}okas}, {Wojtak}, {Gottl{\"o}ber},
  {Mamon}, \& {Prada}}]{lokas06}
{{\L}okas}, E.~L., {Wojtak}, R., {Gottl{\"o}ber}, S., {Mamon}, G.~A., \&
  {Prada}, F. 2006, \mnras, 367, 1463

\bibitem[{{Mahdavi} {et~al.}(2005){Mahdavi}, {Trentham}, \&
  {Tully}}]{mahdavi05}
{Mahdavi}, A., {Trentham}, N., \& {Tully}, R.~B. 2005, \aj, 130, 1502

\bibitem[{{Majumdar} \& {Mohr}(2004)}]{majumdar04}
{Majumdar}, S. \& {Mohr}, J.~J. 2004, \apj, 613, 41

\bibitem[{{Markevitch} {et~al.}(1999){Markevitch}, {Vikhlinin}, {Forman}, \&
  {Sarazin}}]{mvfs}
{Markevitch}, M., {Vikhlinin}, A., {Forman}, W.~R., \& {Sarazin}, C.~L. 1999,
  \apj, 527, 545

\bibitem[{{Markevitch} {et~al.}(2000)}]{2000ApJ...541..542M}
{Markevitch}, M. {et~al.} 2000, \apj, 541, 542

\bibitem[{{Mart{\'{\i}}nez} {et~al.}(2002){Mart{\'{\i}}nez}, {Zandivarez},
  {Merch{\'a}n}, \& {Dom{\'{\i}}nguez}}]{martinez02b}
{Mart{\'{\i}}nez}, H.~J., {Zandivarez}, A., {Merch{\'a}n}, M.~E., \&
  {Dom{\'{\i}}nguez}, M.~J.~L. 2002, \mnras, 337, 1441

\bibitem[{{Massey} {et~al.}(2005){Massey}, {Refregier}, {Bacon}, {Ellis}, \&
  {Brown}}]{massey05}
{Massey}, R., {Refregier}, A., {Bacon}, D.~J., {Ellis}, R., \& {Brown}, M.~L.
  2005, \mnras, 359, 1277

\bibitem[{{Mazure} {et~al.}(1996)}]{mazure96}
{Mazure}, A. {et~al.} 1996, \aap, 310, 31

\bibitem[{{McNamara} {et~al.}(2000){McNamara}, {Wise}, {Nulsen}, {David},
  {Sarazin}, {Bautz}, {Markevitch}, {Vikhlinin}, {Forman}, {Jones}, \&
  {Harris}}]{2000ApJ...534L.135M}
{McNamara}, B.~R., {Wise}, M., {Nulsen}, P. E.~J., {David}, L.~P., {Sarazin},
  C.~L., {Bautz}, M., {Markevitch}, M., {Vikhlinin}, A., {Forman}, W.~R.,
  {Jones}, C., \& {Harris}, D.~E. 2000, \apjl, 534, L135

\bibitem[{{Metzler} {et~al.}(2001){Metzler}, {White}, \&
  {Loken}}]{2001ApJ...547..560M}
{Metzler}, C.~A., {White}, M., \& {Loken}, C. 2001, \apj, 547, 560

\bibitem[{{Miller} {et~al.}(2005)}]{miller05}
{Miller}, C.~J. {et~al.} 2005, \aj, 130, 968

\bibitem[{{Mitchell} {et~al.}(2005){Mitchell}, {Keeton}, {Frieman}, \&
  {Sheth}}]{mitchell05}
{Mitchell}, J.~L., {Keeton}, C.~R., {Frieman}, J.~A., \& {Sheth}, R.~K. 2005,
  \apj, 622, 81

\bibitem[{{Natarajan} {et~al.}(2002){Natarajan}, {Loeb}, {Kneib}, \&
  {Smail}}]{natarajan02}
{Natarajan}, P., {Loeb}, A., {Kneib}, J.-P., \& {Smail}, I. 2002, \apjl, 580,
  L17

\bibitem[{{Navarro} {et~al.}(1997){Navarro}, {Frenk}, \& {White}}]{nfw97}
{Navarro}, J.~F., {Frenk}, C.~S., \& {White}, S. D.~M. 1997, \apj, 490, 493

\bibitem[{{Pierpaoli} {et~al.}(2003){Pierpaoli}, {Borgani}, {Scott}, \&
  {White}}]{pierpaoli03}
{Pierpaoli}, E., {Borgani}, S., {Scott}, D., \& {White}, M. 2003, \mnras, 342,
  163

\bibitem[{{Pierpaoli} {et~al.}(2001){Pierpaoli}, {Scott}, \&
  {White}}]{pierpaoli01}
{Pierpaoli}, E., {Scott}, D., \& {White}, M. 2001, \mnras, 325, 77

\bibitem[{{Pike} \& {Hudson}(2005)}]{pike05}
{Pike}, R.~W. \& {Hudson}, M.~J. 2005, \apj, 635, 11

\bibitem[{{Pisani} {et~al.}(2003){Pisani}, {Ramella}, \& {Geller}}]{pisani03}
{Pisani}, A., {Ramella}, M., \& {Geller}, M.~J. 2003, \aj, 126, 1677

\bibitem[{{Popesso} {et~al.}(2006){Popesso}, {Biviano}, {B{\"o}hringer}, \&
  {Romaniello}}]{popesso06}
{Popesso}, P., {Biviano}, A., {B{\"o}hringer}, H., \& {Romaniello}, M. 2006,
  ArXiv Astrophysics e-prints

\bibitem[{{Popesso} {et~al.}(2005){Popesso}, {Biviano}, {B{\"o}hringer},
  {Romaniello}, \& {Voges}}]{popesso05}
{Popesso}, P., {Biviano}, A., {B{\"o}hringer}, H., {Romaniello}, M., \&
  {Voges}, W. 2005, \aap, 433, 431

\bibitem[{{Popesso} {et~al.}(2004){Popesso}, {B{\"o}hringer}, {Brinkmann},
  {Voges}, \& {York}}]{popesso04}
{Popesso}, P., {B{\"o}hringer}, H., {Brinkmann}, J., {Voges}, W., \& {York},
  D.~G. 2004, \aap, 423, 449

\bibitem[{{Press} \& {Schechter}(1974)}]{ps74}
{Press}, W.~H. \& {Schechter}, P. 1974, \apj, 187, 425

\bibitem[{{Rasia} {et~al.}(2005){Rasia}, {Mazzotta}, {Borgani}, {Moscardini},
  {Dolag}, {Tormen}, {Diaferio}, \& {Murante}}]{rasia05}
{Rasia}, E., {Mazzotta}, P., {Borgani}, S., {Moscardini}, L., {Dolag}, K.,
  {Tormen}, G., {Diaferio}, A., \& {Murante}, G. 2005, \apjl, 618, L1

\bibitem[{{Reiprich}(2006)}]{reiprich06}
{Reiprich}, T.~H. 2006, ArXiv Astrophysics e-prints

\bibitem[{{Reiprich} \& {B{\" o}hringer}(2002)}]{hiflugcs}
{Reiprich}, T.~H. \& {B{\" o}hringer}, H. 2002, \apj, 567, 716

\bibitem[{{Reisenegger} {et~al.}(2000){Reisenegger}, {Quintana}, {Carrasco}, \&
  {Maze}}]{rqcm}
{Reisenegger}, A., {Quintana}, H., {Carrasco}, E.~R., \& {Maze}, J. 2000, \aj,
  120, 523

\bibitem[{{Rines} {et~al.}(2004){Rines}, {Geller}, {Diaferio}, {Kurtz}, \&
  {Jarrett}}]{cairnsii}
{Rines}, K., {Geller}, M.~J., {Diaferio}, A., {Kurtz}, M.~J., \& {Jarrett},
  T.~H. 2004, \aj, 128, 1078

\bibitem[{{Rines} {et~al.}(2002){Rines}, {Geller}, {Diaferio}, {Mahdavi},
  {Mohr}, \& {Wegner}}]{rines02}
{Rines}, K., {Geller}, M.~J., {Diaferio}, A., {Mahdavi}, A., {Mohr}, J.~J., \&
  {Wegner}, G. 2002, \aj, 124, 1266

\bibitem[{{Rines} {et~al.}(2000){Rines}, {Geller}, {Diaferio}, {Mohr}, \&
  {Wegner}}]{rines2000}
{Rines}, K., {Geller}, M.~J., {Diaferio}, A., {Mohr}, J.~J., \& {Wegner}, G.~A.
  2000, \aj, 120, 2338

\bibitem[{{Rines} {et~al.}(2003){Rines}, {Geller}, {Kurtz}, \&
  {Diaferio}}]{cairnsi}
{Rines}, K., {Geller}, M.~J., {Kurtz}, M.~J., \& {Diaferio}, A. 2003, \aj, 126,
  2152

\bibitem[{{Rines} {et~al.}(2005){Rines}, {Geller}, {Kurtz}, \&
  {Diaferio}}]{cairnsha}
---. 2005, \aj, 130, 1482

\bibitem[{{Rines} {et~al.}(2001){Rines}, {Geller}, {Kurtz}, {Diaferio},
  {Jarrett}, \& {Huchra}}]{rines01a}
{Rines}, K., {Geller}, M.~J., {Kurtz}, M.~J., {Diaferio}, A., {Jarrett}, T.~H.,
  \& {Huchra}, J.~P. 2001, \apj, 561, L41

\bibitem[{{Rines} \& {Diaferio}(2006)}]{cirsi}
{Rines}, K.~J. \& {Diaferio}, A. 2006, \aj, in press (astro-ph/0602032)

\bibitem[{{Schmidt}(1968)}]{schmidt68}
{Schmidt}, M. 1968, \apj, 151, 393

\bibitem[{{Schuecker} {et~al.}(2003){Schuecker}, {B{\" o}hringer}, {Collins},
  \& {Guzzo}}]{schuecker03}
{Schuecker}, P., {B{\" o}hringer}, H., {Collins}, C.~A., \& {Guzzo}, L. 2003,
  \aap, 398, 867

\bibitem[{{Seljak}(2002)}]{seljak02}
{Seljak}, U. 2002, \mnras, 337, 769

\bibitem[{{Seljak} {et~al.}(2006){Seljak}, {Slosar}, \& {McDonald}}]{seljak06}
{Seljak}, U., {Slosar}, A., \& {McDonald}, P. 2006, ArXiv Astrophysics e-prints

\bibitem[{{Semboloni} {et~al.}(2006){Semboloni}, {Mellier}, {van Waerbeke},
  {Hoekstra}, {Tereno}, {Benabed}, {Gwyn}, {Fu}, {Hudson}, {Maoli}, \&
  {Parker}}]{semboloni06}
{Semboloni}, E., {Mellier}, Y., {van Waerbeke}, L., {Hoekstra}, H., {Tereno},
  I., {Benabed}, K., {Gwyn}, S.~D.~J., {Fu}, L., {Hudson}, M.~J., {Maoli}, R.,
  \& {Parker}, L.~C. 2006, \aap, 452, 51

\bibitem[{{Sheldon} {et~al.}(2001)}]{sheldon01}
{Sheldon}, E.~S. {et~al.} 2001, \apj, 554, 881

\bibitem[{{Sheth} \& {Diaferio}(2001)}]{sheth01}
{Sheth}, R.~K. \& {Diaferio}, A. 2001, \mnras, 322, 901

\bibitem[{{Sheth} \& {Tormen}(1999)}]{sheth99}
{Sheth}, R.~K. \& {Tormen}, G. 1999, \mnras, 308, 119

\bibitem[{{Sheth} {et~al.}(2003)}]{sheth03}
{Sheth}, R.~K. {et~al.} 2003, \apj, 594, 225

\bibitem[{{Shimasaku}(1993)}]{shimasaku93}
{Shimasaku}, K. 1993, \apj, 413, 59

\bibitem[{{Shimasaku}(1997)}]{shimasaku97}
---. 1997, \apj, 489, 501

\bibitem[{{Shimizu} {et~al.}(2006){Shimizu}, {Kitayama}, {Sasaki}, \&
  {Suto}}]{shimizu06}
{Shimizu}, M., {Kitayama}, T., {Sasaki}, S., \& {Suto}, Y. 2006, \pasj, 58, 291

\bibitem[{{Slosar} {et~al.}(2006){Slosar}, {Seljak}, \&
  {Tasitsiomi}}]{2006MNRAS.366.1455S}
{Slosar}, A., {Seljak}, U., \& {Tasitsiomi}, A. 2006, \mnras, 366, 1455

\bibitem[{{Spergel} {et~al.}(2003)}]{spergel03}
{Spergel}, D. {et~al.} 2003, \apjs, 148, 175

\bibitem[{{Spergel} {et~al.}(2006)}]{spergel06}
{Spergel}, D.~N. {et~al.} 2006, ArXiv Astrophysics e-prints

\bibitem[{{Stanek} {et~al.}(2006){Stanek}, {Evrard}, {B{\"o}hringer},
  {Schuecker}, \& {Nord}}]{stanek06}
{Stanek}, R., {Evrard}, A.~E., {B{\"o}hringer}, H., {Schuecker}, P., \& {Nord},
  B. 2006, ArXiv Astrophysics e-prints

\bibitem[{{Stoughton} {et~al.}(2002)}]{sdss}
{Stoughton}, C. {et~al.} 2002, \aj, 123, 485

\bibitem[{{Strauss} {et~al.}(2002)}]{strauss02}
{Strauss}, M.~A. {et~al.} 2002, \aj, 124, 1810

\bibitem[{{Sugiyama}(1995)}]{sugiyama95}
{Sugiyama}, N. 1995, \apjs, 100, 281

\bibitem[{{Tinker} {et~al.}(2005){Tinker}, {Weinberg}, {Zheng}, \&
  {Zehavi}}]{tinker05}
{Tinker}, J.~L., {Weinberg}, D.~H., {Zheng}, Z., \& {Zehavi}, I. 2005, \apj,
  631, 41

\bibitem[{{van den Bosch} {et~al.}(2003){van den Bosch}, {Mo}, \&
  {Yang}}]{2003MNRAS.345..923V}
{van den Bosch}, F.~C., {Mo}, H.~J., \& {Yang}, X. 2003, \mnras, 345, 923

\bibitem[{{van den Bosch} {et~al.}(2005){van den Bosch}, {Weinmann}, {Yang},
  {Mo}, {Li}, \& {Jing}}]{2005MNRAS.361.1203V}
{van den Bosch}, F.~C., {Weinmann}, S.~M., {Yang}, X., {Mo}, H.~J., {Li}, C.,
  \& {Jing}, Y.~P. 2005, \mnras, 361, 1203

\bibitem[{{van der Marel} {et~al.}(2000){van der Marel}, {Magorrian},
  {Carlberg}, {Yee}, \& {Ellingson}}]{2000AJ....119.2038V}
{van der Marel}, R.~P., {Magorrian}, J., {Carlberg}, R.~G., {Yee}, H.~K.~C., \&
  {Ellingson}, E. 2000, \aj, 119, 2038

\bibitem[{{Viana} {et~al.}(2003){Viana}, {Kay}, {Liddle}, {Muanwong}, \&
  {Thomas}}]{viana03}
{Viana}, P.~T.~P., {Kay}, S.~T., {Liddle}, A.~R., {Muanwong}, O., \& {Thomas},
  P.~A. 2003, \mnras, 346, 319

\bibitem[{{Viana} {et~al.}(2002){Viana}, {Nichol}, \& {Liddle}}]{viana02}
{Viana}, P.~T.~P., {Nichol}, R.~C., \& {Liddle}, A.~R. 2002, \apjl, 569, L75

\bibitem[{{Vikhlinin} {et~al.}(2006){Vikhlinin}, {Kravtsov}, {Forman}, {Jones},
  {Markevitch}, {Murray}, \& {Van Speybroeck}}]{vikhlinin06}
{Vikhlinin}, A., {Kravtsov}, A., {Forman}, W., {Jones}, C., {Markevitch}, M.,
  {Murray}, S.~S., \& {Van Speybroeck}, L. 2006, \apj, 640, 691

\bibitem[{{Vikhlinin} {et~al.}(2001){Vikhlinin}, {Markevitch}, \&
  {Murray}}]{2001ApJ...551..160V}
{Vikhlinin}, A., {Markevitch}, M., \& {Murray}, S.~S. 2001, \apj, 551, 160

\bibitem[{{Vikhlinin} {et~al.}(2003)}]{vikhlinin03}
{Vikhlinin}, A. {et~al.} 2003, \apj, 590, 15

\bibitem[{{Voevodkin} \& {Vikhlinin}(2004)}]{vv04}
{Voevodkin}, A. \& {Vikhlinin}, A. 2004, \apj, 601, 610

\bibitem[{{Voges} {et~al.}(1999)}]{rass}
{Voges}, W. {et~al.} 1999, \aap, 349, 389

\bibitem[{{Warren} {et~al.}(2005){Warren}, {Abazajian}, {Holz}, \&
  {Teodoro}}]{warren06}
{Warren}, M.~S., {Abazajian}, K., {Holz}, D.~E., \& {Teodoro}, L. 2005, ArXiv
  Astrophysics e-prints

\bibitem[{{White}(2002)}]{white02}
{White}, M. 2002, \apjs, 143, 241

\bibitem[{{Yoshikawa} {et~al.}(2003){Yoshikawa}, {Jing}, \&
  {B{\"o}rner}}]{2003ApJ...590..654Y}
{Yoshikawa}, K., {Jing}, Y.~P., \& {B{\"o}rner}, G. 2003, \apj, 590, 654

\bibitem[{{Zabludoff} {et~al.}(1993){Zabludoff}, {Geller}, {Huchra}, \&
  {Ramella}}]{1993AJ....106.1301Z}
{Zabludoff}, A.~I., {Geller}, M.~J., {Huchra}, J.~P., \& {Ramella}, M. 1993,
  \aj, 106, 1301

\bibitem[{{Zwicky}(1933)}]{zwicky1933}
{Zwicky}, F. 1933, Helv.~Phys.~Acta, 6, 110

\bibitem[{{Zwicky}(1937)}]{zwicky1937}
---. 1937, \apj, 86, 217

\end{thebibliography}

\clearpage

\epsscale{1}

\epsscale{1}

\epsscale{1}

\epsscale{1}

\epsscale{1}

\epsscale{1}

\epsscale{1}

\epsscale{1}

\end{document}